%% file: barprop10pp.tex
\documentclass{article}

\usepackage{emulateapj}
\usepackage{psfig}

\lefthead{Weiner \& Sellwood}
\righthead{Properties of the Galactic Bar}

\slugcomment{\it Submitted to ApJ}

\begin{document}

\long\def\Ignore#1{\relax}

\long\def\Comment#1{\relax}

\def\cp{\clearpage}

\newcommand\kms{\hbox{$\rm{km}~\rm{s}^{-1}$}}
\newcommand\ha{\hbox{H$\alpha$}}
\newcommand\hi{\hbox{H~I}} 
\newcommand\el{\hbox{$\ell$}}
\newcommand\philsr{\hbox{$\phi_{\rm LSR}$}} 
\newcommand\lv{$\ell-V\,$}
\newcommand\mden{\hbox{${\rm M}_{\odot}~{\rm pc^{-3}}$}}
\newcommand\msurfden{\hbox{${\rm M}_{\odot}~{\rm pc^{-2}}$}}
\newcommand\kmskpc{\hbox{$\rm{km}~\rm{s}^{-1}~\rm{kpc}^{-1}$}}
\def\deg{\hbox{$^{\circ}$}} 
\newcommand\msun{\hbox{${\rm M}_{\odot}$}}
\newcommand\degk{\hbox{ K}} 
\newcommand\lvunit{$\msun\,{\rm kpc}^{-2}\,{\rm deg}^{-1}\,/\,\kms$} 
\newcommand\etal{{\it et al.}}
\newcommand\eg{{\it e.g.\,}}


\title{The Properties of the Galactic Bar Implied by \\ 
Gas Kinematics in the Inner Milky Way}

\author{Benjamin J.\ Weiner\altaffilmark{1,2} 
and J.\ A.\ Sellwood\altaffilmark{2}}

\affil{$^1$Observatories of the Carnegie Institution of Washington, \\ 
813 Santa Barbara St, Pasadena, CA  91101}

\affil{$^2$Department of Physics and Astronomy, Rutgers, The State
University \\ 136 Frelinghuysen Road, Piscataway, NJ 08854-8019}

\affil{Electronic mail: bjw@ociw.edu, sellwood@physics.rutgers.edu}

\quad\newline \centerline{Rutgers Astrophysics Preprint No.\ 236}

\begin{abstract}

Longitude-velocity ($\ell-V$) diagrams of H~I and CO gas in the inner Milky
Way have long been known to be inconsistent with circular motion in an
axisymmetric potential.  Several lines of evidence suggest that the Galaxy is
barred, and gas flow in a barred potential could be consistent with
the observed ``forbidden'' velocities and other features in the data.
We compare the H~I observations to \lv
diagrams synthesized from 2--D fluid dynamical simulations of gas
flows in a family of barred potentials.  The gas flow pattern is very
sensitive to the parameters of the assumed potential, which allows us
to discriminate among models.  We present a model that reproduces the
outer contour of the \hi\ \lv diagram reasonably well; this model has
a strong bar with a semimajor axis of 3.6 kpc, an axis ratio 
of approximately 3:1, an inner Lindblad resonance (ILR), 
and a pattern speed of 42 \kmskpc, and matches the data 
best when viewed from 34\deg\ to the bar major axis.
The behavior of the models, combined with the constraint that the
shocks in the Milky Way bar should resemble those in external barred
galaxies, leads us to conclude that wide ranges of parameter space are
incompatible with the observations.  In particular we suggest that the
bar must be fairly strong, must have an ILR, and cannot be too end-on,
with the bar major axis at $35\deg \pm 5\deg$ to the line of sight.
The \hi\ data exhibit larger forbidden velocities over a wider longitude
range than are seen in molecular gas; this important difference is the
reason our favored model differs so significantly from other recently
proposed models.

\end{abstract}


\section{Introduction}

The structure and morphology of the inner Milky Way are difficult to
determine due both to dust obscuration and to our edge-on view. 
The canonical picture of the Milky Way as an
axisymmetric spiral galaxy was enshrined in the models of Schmidt
(1965), Bahcall \& Soneira (1980), Ostriker \& Caldwell (1983), Kent
(1992), and others.  However, the suggestion by de Vaucouleurs (1964)
that the Galaxy is barred has been supported by many recent studies
(cf.\ the reviews of Blitz \etal\ 1993 and Kuijken 1996).  What was
once thought of as the bulge now seems to be, at least in part, a
thickened bar.  Lines of evidence for a bar
include: the infrared surface brightness
distribution (Blitz \& Spergel 1991; Dwek \etal\ 1995), the
distribution of Mira variables (Whitelock \& Catchpole 1992), IRAS
point sources (Weinberg 1992, Nikolaev \& Weinberg 1997), 
the magnitude offset of bulge stars at
positive and negative longitudes (Stanek 1995; Stanek \etal\ 1997),
OH/IR stars (Sevenster 1995),
and the gas motions near the Galactic center (\eg\ Liszt \& Burton 
1980; Binney \etal\ 1991).  Several groups have
used infrared photometry, especially from the COBE/DIRBE data, to
deduce the density distribution in the Galactic bar (\eg\ Blitz \&
Spergel 1991; Dwek \etal\ 1995; Binney, Gerhard \& Spergel 1997).

It has long been known, from both 21~cm and mm observations of gaseous
emission lines, that the kinematics of gas toward the Galactic center
($|l| \lesssim 10\deg$) are inconsistent with purely circular motions
(\eg\ Rougoor \& Oort 1960, Kerr \& Westerhout 1965, Oort 1977).
Figure \ref{fig-lisztdata} shows the \hi\ longitude--velocity
($\ell-V$) diagram constructed from the data of Liszt \& Burton
(1980; see also Burton \& Liszt 1983).  This diagram shows the
distribution of \hi\ radial velocities at galactic longitudes 
$13\deg > \ell > -11\deg$.  
Most gas is approaching at negative longitudes and
receding at positive, which is the general sense of rotation of the
Milky Way, but there is significant emission from gas moving in the
opposite sense on both sides; such gas is inconsistent with simple
circular orbits and is said to have ``forbidden velocities.''
Forbidden velocities in excess of 100 \kms\ are observed throughout
the range $-6\deg < \ell < 6\deg$.  

A variety of explanations for the non-circular motions have been proposed
including explosive outflows (cf.\ Oort 1977),
spiral density waves (\eg\ Scoville, Solomon \& Jefferts 1974),
and barlike perturbations.
If the non-circular motions do result from gas flow in a non-axisymmetric
potential, observation and detailed modeling of the gas kinematics
should provide strong constraints on the mass distribution in the inner
Galaxy.  In fact, flow patterns in barred galaxy models have already
been shown to provide qualitative fits to the observations (\eg\ 
Peters 1975; Liszt \& Burton 1980; van Albada 1985b; Mulder \& 
Liem 1986; Binney \etal\ 1991).

Features in diagrams such as Figure \ref{fig-lisztdata} contain
information about the
distribution of gas in space and velocity within the disk of the
Galaxy.  But because we cannot determine the distance to individual
parcels of gas, there is no unique way to invert the observed \lv
diagram to determine the two-dimensional distribution of gas in the
Galaxy; the projection into $\ell$ and $V$ space is highly degenerate.
Even if such a deprojection were available, we still could not use the
flow pattern to deduce the galactic gravitational potential directly,
since the gas is also subject to pressure forces and its motion is
governed by the non-linear equations of fluid dynamics.

Thus the data need to be interpreted by comparison with models.
Binney \etal\ (1991) compare stellar orbits in a barred model with the
CO and \hi\ \lv diagrams, which offers some insight, but omits the
effects of the strong shocks expected in gas flows in a bar.  Subsequently,
several numerical methods have been employed to construct improved
models for the gas.  Jenkins \& Binney (1994) used sticky particles,
Englmaier \& Gerhard (1998) used smoothed particle hydrodynamics (SPH), 
while Fux (1997,1999) combined SPH
and $N$-body techniques to attempt to build a fully self-consistent
model of the inner Milky Way.  Fux (1999) has compared the gas kinematics
in such a model to ``arm'' features in the CO and \hi\ \lv diagrams 
to constrain the properties of the bar; his approach is complementary
to ours, concentrating on high-density regions of the \lv diagram.

Most modeling efforts have been devoted to observations of
the dense molecular gas while comparatively little attention
has been devoted to the \hi\ data.  Here we focus on the \lv diagram
for the \hi, which is less affected by two principal limitations
of the molecular data: the \lv diagram for the \hi\ is both more
symmetric and more complete than the corresponding CO plots.  
In particular, CO (Dame
\etal\ 1987; Bally \etal\ 1988) is not detected where \hi\
emission is present in some significant regions of the \lv
plane; for example, between $\ell = 0\deg$ and
$-6\deg$, the \hi\ emission extends to $\sim -270~\kms$ while the
CO emission extends to $\sim -220~\kms$ only (Figure 4 of
Dame \etal).
More importantly, \hi\ emission extends to higher forbidden velocities 
over a wider angular range in comparison with that observed in CO.

We attempt to place constraints on the properties of the Galactic bar
by comparing the \hi\ \lv diagram with similar plots synthesized from
many fluid-dynamical models in various potentials.  The full gas
velocity field allows us to determine which regions of the Galaxy are
responsible for prominent features of the \lv diagram.  Our goal is
not to identify a unique model, but rather to infer properties of the
inner Galaxy that appear to be required by the data. 
We conclude that the Galaxy must have a strong bar that rotates fairly 
quickly and has a central density high enough to
produce an inner Lindblad resonance.  The bar must have a
semi-major axis $a \gtrsim 3$~kpc, and be
viewed obliquely, with the bar major axis between 30\deg\ and 40\deg\
to the Sun--Galactic Center line.

\section{The Galactic longitude--velocity diagram}
\label{sec-lvdiag}

\subsection{Observational data}

\begin{figure*}[t]

\psfig{figure=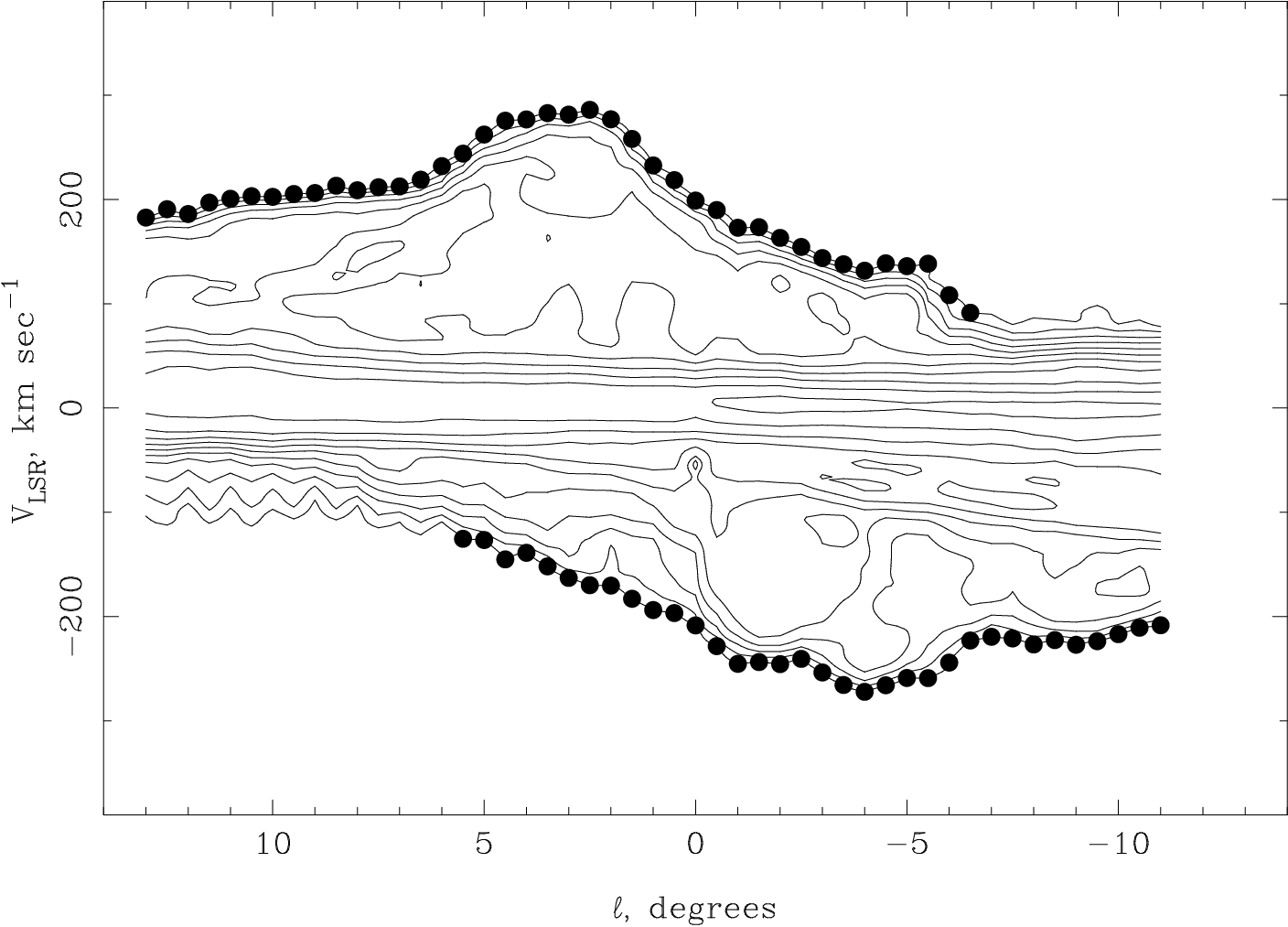,width=7.0truein,angle=-90}

\caption{The longitude--velocity diagram of \hi\ in the inner Galaxy
from the data of Liszt \& Burton (1980).  The lowest contour is
0.125\degk\ in antenna temperature summed over $b$, or 1.25 \lvunit\
of atomic gas (H+He), and the contours increase geometrically by a
factor of 2.  The forbidden quadrants are positive $V$ at negative
\el, and negative $V$ at positive \el.  The band of emission at $|V|
\lesssim 100~\kms$ is foreground from the disk.  The filled circles
are those data points on the extreme-velocity contour that we try to
reproduce in the simulations.
\label{fig-lisztdata}
}

\end{figure*}

We use the \hi\ observations of the inner Galaxy by Burton \& Liszt
(1978, 1983; and Liszt \& Burton 1980), which produced the 
\lv diagram shown in Figure \ref{fig-lisztdata}.
These data have
uniform coverage of the longitude range $\ell = -11\deg$ to +13\deg,
with spatial resolution $\sim 0.5\deg$, well matched to the resolution
of our simulations, and good velocity resolution ($2.75~\kms$)
and sensitivity.  H.\ Liszt kindly provided the data in electronic form.  
The spectra are
taken on an 0.5\deg\ grid in \el\ and $b$; because we are comparing to
2-D simulations, we summed the data along the $b$ axis.  We also
smoothed in $V$ with a Gaussian of $\sigma = 5.5~\kms$.
A high-velocity \hi\ cloud at $\ell=8\deg, b=-4\deg$ and $V=-210~\kms$
(``Shane's feature,'' Saraber \& Shane 1974),
was excluded from the dataset.

\placefigure{fig-lisztdata}

Plots of individual latitude slices (Burton \& Liszt 1978; 
Liszt \& Burton 1980) show that
the velocity ``peaks'' in Figure \ref{fig-lisztdata} are prominent at
latitudes near $b = 0\deg$.  The broad band of emission 
(sometimes called the ``main maximum'') at $-100
< V < 100\ \kms$ at all longitudes is present over the entire latitude
range observed by Liszt \& Burton ($-6\deg < b < 6\deg$).  Since the
half-thickness of the gas layer is approximately 250 pc inside the
Solar radius to 4 kpc radius, and the thickness may be only 100 kpc
inside 4 kpc (Mihalas \& Binney 1981; Jackson \& Kellman 1974), the
band of emission is presumably from disk gas that is relatively close
by.  The velocity extent of the band is large for the velocity dispersion 
of the gas as derived by Gunn, Knapp \& Tremaine (1979),
even given the 1000:1 density contrast.  It is presumably 
attributable to line-of-sight integration over
substantial bulk motions in the disk such as spiral arm streaming motions
(Burton \& Liszt 1983).

Foreground gas is also responsible for 21 cm absorption against the
central continuum source at $\ell=0\deg$, $b=0\deg$.  This 
absorption appears at negative velocity (Burton \& Liszt 1978, 1993)
and is visible in the summed data in some of the intermediate 
contours in Figure~\ref{fig-lisztdata}, although it is not
conspicuous in the extreme contour.  The absorption at negative
velocities implies that the negative-velocity gas at $\ell=0\deg$, $b=0\deg$
is between the Sun and the Galactic Center, while the positive
velocity gas at that position is behind the Center (Burton \& Liszt 1978).

The filled circles in Figure \ref{fig-lisztdata} mark the points of
the observed extreme-velocity contour (EVC) we will use for 
comparison to the simulations.
Because we are interested in the motions of the gas in the inner
Galaxy, we do not use that portion of the EVC that appears to be
substantially influenced by foreground disk gas, but we retain the
data point at $\ell=0\deg$ since the extreme contour there is not
much affected by absorption.

The \lv diagram is not perfectly two-fold
rotationally symmetric in many respects.  Here we simply note
that the shapes of the velocity peaks
in the EVC differ: that at positive \el\ lies at 3\deg\
while the most negative velocity is at $\ell =-4\deg$, although the
magnitudes are similar.  More detailed plots of the \hi\ \lv
diagram reveal other non-symmetric features in the interior
of the diagram, including the well-known ``3-kpc expanding arm'' 
(\eg\ Peters 1975; Burton \& Liszt 1983, 1993), which is
marginally visible in Figure~\ref{fig-lisztdata} at 
$-3\deg > \ell > -9\deg$ near $-100~\kms$.  
Some investigators (\eg\ Kerr 1967; Liszt \& Burton 1980)
have also presented evidence that the \hi\ gas
distribution in the inner Galaxy is tilted out of the Galactic
plane.  By summing the data over $b$, we have suppressed
this aspect, which would be difficult to address in any case
since our models are two-dimensional.

\subsection{Interpreting the extreme-velocity contour} 

One must make assumptions in order to extract information on the
structure of the Galaxy from the \lv diagram.  The simplest approach
is to assume that the Galaxy is axisymmetric and the gas moves on
circular orbits.

With this assumption (and others noted below), the \lv diagram can be
used to determine the rotation curve of the Galaxy interior to the
Solar circle by the tangent-point method
(cf.\ Gunn \etal\ 1979, Mihalas \& Binney
1981).  The critical feature of the \lv diagram in this method is the
extreme-velocity contour (EVC), which is the outer contour of the gas
distribution in longitude--velocity space; it is the highest absolute
radial velocity observed along the line of sight at each \el.  In
the tangent-point method, 
the EVC in the upper left and lower right quadrants only is used; gas
at forbidden velocities is ignored.  
The extreme observed velocity needs to be corrected for instrumental
resolution and the velocity dispersion of the gas, which is assumed to
have a uniform value (Gunn \etal\ 1979), to find the terminal velocity
at each longitude, $v_t(\ell)$ (see Section \ref{sec-evclevel} below).
With the further assumptions that some \hi\
gas exists at every tangent point and that the circular angular frequency,
$\Omega(R)$, decreases monotonically from the center, $v_t(\ell)$
yields the Galactic rotation curve $\Theta(R)$ directly through the
equation 
$\Theta(R_0~|\rm{sin}~\ell |) = |v_t(\ell)| + \Theta_0~|\rm{sin}~\ell|$.

As the correction term for the circular velocity of the LSR is small 
at longitudes near $0\deg$, the EVC on the maximum side
(positive $V$ at positive \el, negative $V$ at negative \el) is
approximately the rotation curve under the axisymmetric assumption.  

For circular orbits, on one side of the Galactic center all
the gas should be coming towards the Sun, and on the other side it
should be going away.  Hence, the EVC on the non-maximum side, in
the upper right and lower left quadrants of the \lv diagram, should
be featureless and close to 0 \kms\ (as long as 
the circular frequency at $R_0$ is less than the circular frequency
in the inner Galaxy, which is true for any reasonable rotation curve).
The velocity dispersion of the gas and bulk
motions in the disk will push the EVC beyond 0 \kms, but apart from
these effects the non-maximum EVC should not tell us much. 

\begin{figure*}[t]

\psfig{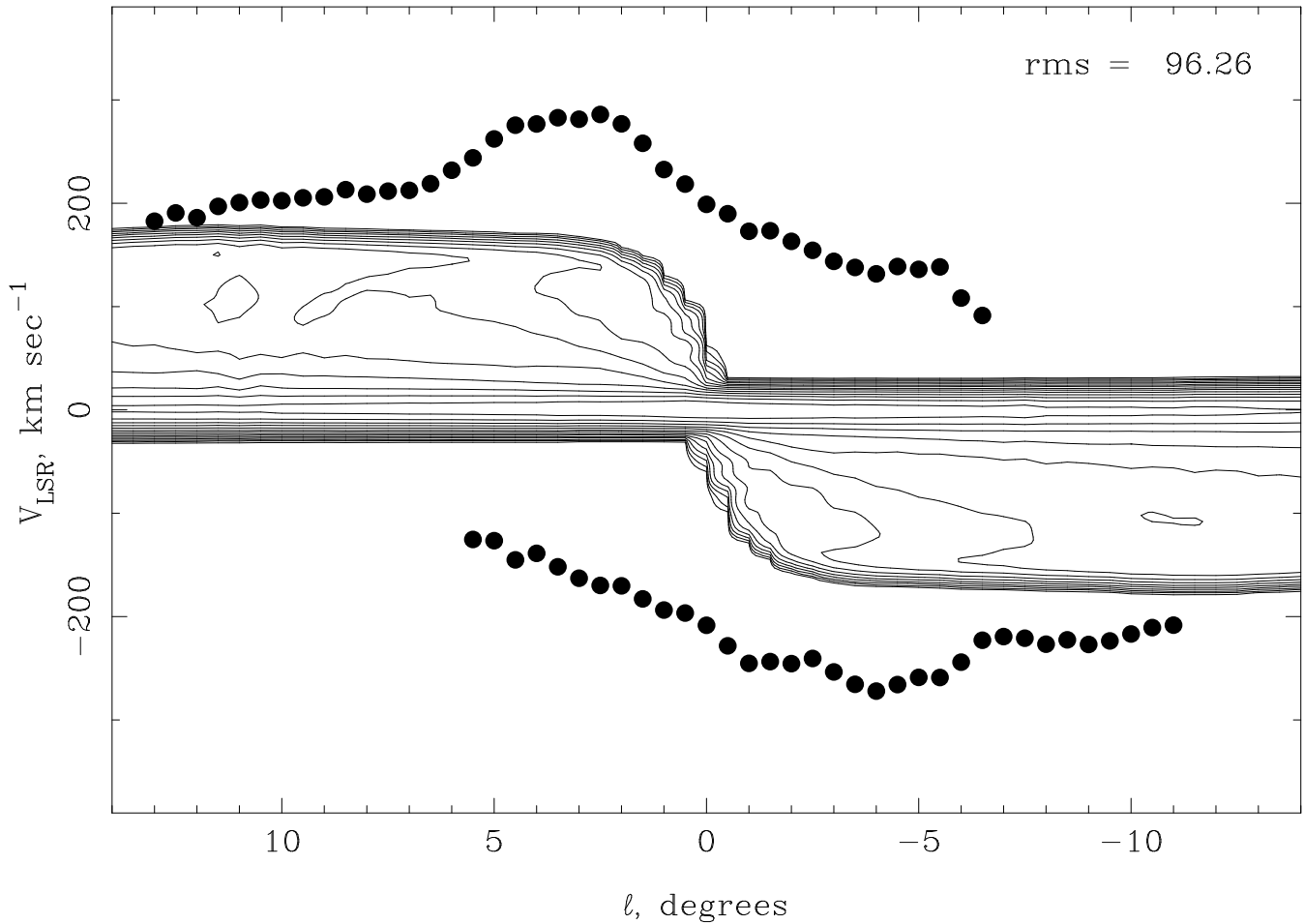}

\caption{The longitude-velocity diagram of gas in an axisymmetric
model.
There is no gas in the forbidden quadrants, beyond the
velocity dispersion-induced spread.  The lowest contour is 1.7
\lvunit\ of total gas.  The contours increase geometrically by a 
factor of 2.
The filled circles are the EVC from the data in Figure
\ref{fig-lisztdata}.  Simply rearranging some of the mass of this model
into a bar yields a model that fits these data points quite well.
\label{fig-axisymlv}
}

\end{figure*}

\placefigure{fig-axisymlv}

\begin{figure*}[t]

\psfig{figure=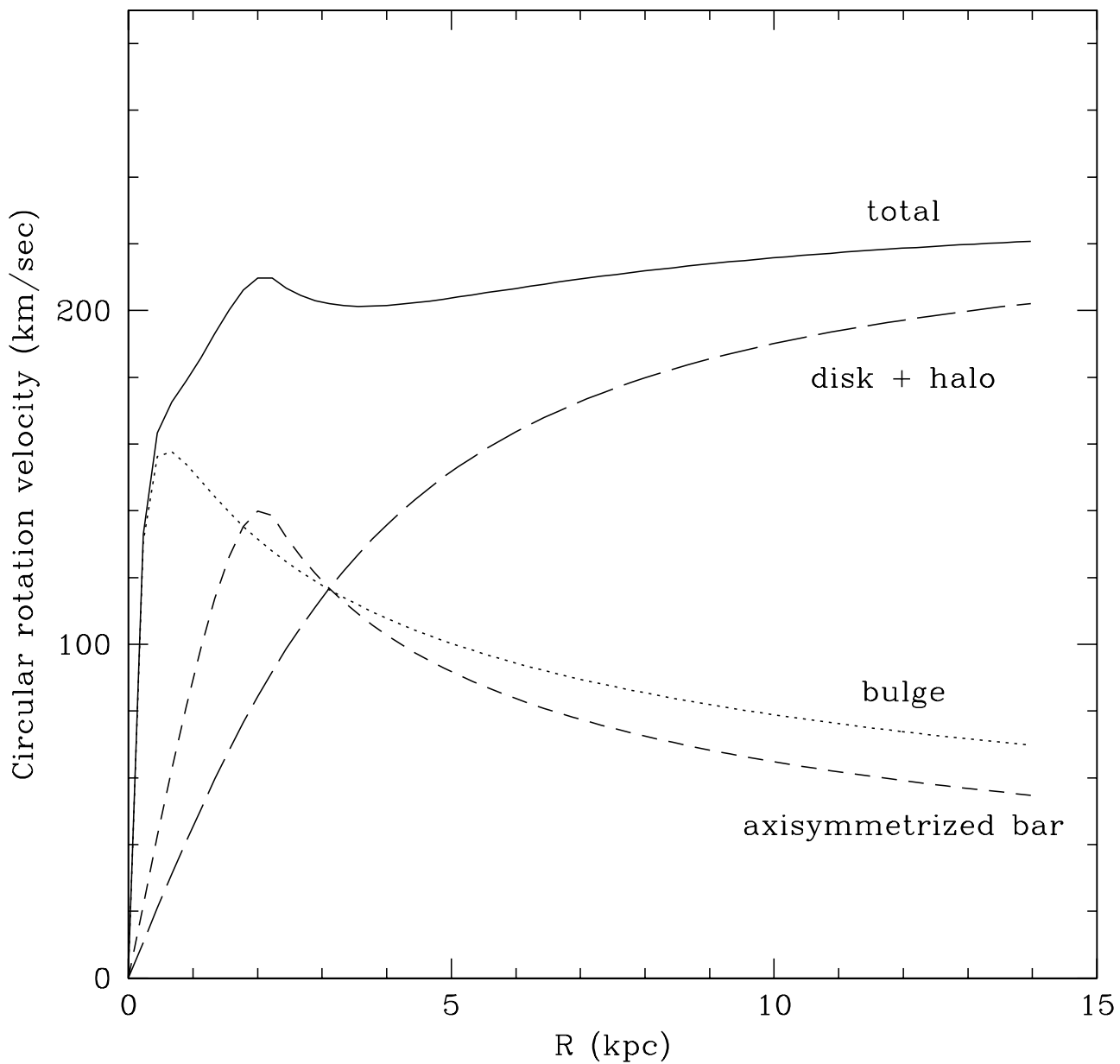,width=7.0truein}

\caption{The rotation curve for the axisymmetric model.  This model is
the axially symmetrized mass distribution of our Model 1.
\label{fig-axirotcurv}
}

\end{figure*}

Figure \ref{fig-axisymlv} shows an \lv diagram for a model with gas
all on circular orbits.  The rotation curve that gives rise to this
\lv diagram is plotted in Figure \ref{fig-axirotcurv}.  The contrast
with Figure \ref{fig-lisztdata} is instructive.  Gas at forbidden
velocities in the Milky Way is clearly inconsistent with a simple
circular flow pattern.

\placefigure{fig-axirotcurv}

The EVC is still a useful probe of the Galactic mass distribution even
when the gas is {\it not\/} on circular orbits, provided
that the observed tracer is ubiquitous in
the disk and that the non-circular motions are caused by streaming in a
non-axisymmetric potential, as first proposed by de Vaucouleurs (1964).  
As long as the observations are sensitive
enough to pick up the tracer in regions of low density, the EVC
depends almost solely on the velocity field, and variations in the
fraction of gas mass in a given tracer phase are much less important.
Here we discount the alternative possibility that non-circular motions
arise from explosions or other violent events near the Galactic Center
(cf.\ Oort 1977).

Neutral hydrogen is ubiquitous in the Galactic disk and is readily
detectable through its 21 cm emission.  It is clearly
more widespread than CO in the inner Galaxy, since there are
no ``holes'' in the \hi\ \lv diagram (Figure \ref{fig-lisztdata}) in
contrast with that for the CO (\eg Figure 4 of Dame \etal\ 1987,
and Figure 4 of Bally \etal\ 1988).  Additionally, as noted earlier, 
the negative velocity peak of the CO \lv diagram reaches only to
$-220~\kms$ at $\ell=-2\deg$ while that peak reaches $-270~\kms$
at $\ell=-4\deg$ in the \hi\ \lv diagram, and the forbidden
emission extends further in \hi\ than in CO, especially 
for negative velocities at $0\deg < \ell < 5\deg$.

The interior of an \lv diagram for CO shows much substructure with strong 
density contrasts, whereas that for \hi\ exhibits only mild variations 
(Figure~\ref{fig-lisztdata}).  Interior features, in both molecular and 
atomic gas, provide extra information to constrain models; \eg\ Fux (1999) 
attempts to match them to an SPH gas flow in a model of the Galaxy.

The additional substructure in molecular emission, which traces gas of 
higher density, is probably caused by variations both in the atomic fraction 
and in molecular emissivity (\eg\ temperature).  Such variations, even if 
they are well understood, would be very hard to model, however.  The 
EVC of the \hi\ \lv diagram, on the other hand, is insensitive to density 
variations.  All successful models of the inner Milky Way should therefore 
match it provided only that there is {\it some\/} atomic gas everywhere in 
the flow.  The smoothness of the EVC in Figure~\ref{fig-lisztdata} gives us 
grounds to hope that this requirement is fulfilled.

\section{Simulations of the gas flow}
\label{sec-simul}

We use a two-dimensional grid-based gas dynamical code to simulate the
gas flow in models for the galactic potential.  The code was
originally written by G.\ D.\ van Albada to model gas flow in barred
galaxy potentials (van Albada 1985a, 1985b) and kindly provided by E.\
Athana\-ssoula.  She used it (Athana\-ssoula 1992b) to study gas flow
patterns in various barred potentials.

\subsection{The fluid code}

The code is an second-order, flux-splitting Eulerian grid code for an
isothermal gas in an imposed gravitational potential representing the
stellar component and halo of the Galaxy.  We neglect the self-gravity
of the gas in order to reduce computational requirements.  We
justify this omission on the grounds that the gas surface density 
is considerably less
than that of the stellar bulge and disk, especially in the inner
regions of the Galaxy with which we are primarily concerned
(see Section~\ref{sec-propmodel} below).

Our grid has 200 by 400 cells, each 50 pc square, and we enforce a
180\deg\ rotation symmetry, so that the grid is effectively 400 by
400.  The grid is fixed with respect to the barred potential,
and both rotate at a steady pattern
speed; the bar is aligned at 45\deg\ to the grid axes.  The time
step is variable, chosen automatically via a Courant condition, and is
generally approximately 0.1 Myr.  The sound speed of the gas is taken
to be 8 \kms\ (cf.\ Gunn \etal\ 1979), corresponding to a temperature
of $\sim 10^4~\degk$.  Varying the sound speed within reasonable limits of
a few \kms\ does not materially affect the derived gas flow.

By its nature, the code approximates the interstellar medium as an
Eulerian fluid, smooth on scales of the grid cell size.  Without some
idealization it is hopeless to simulate the extremely complex dynamics
of the multiphase ISM, which has structure on all observed scales and
a vast assortment of energy inputs and outputs.  Some authors (Jenkins
\& Binney 1994; Combes 1996) have suggested that smooth-fluid models
using the Euler equations, such as grid codes and smooth-particle
hydrodynamics, are not appropriate for the clumpy ISM, and have
advocated various sticky-particle methods.  Sticky-particle codes may
be well suited to simulating the dynamics of the self-gravitating
molecular cloud component, which Jenkins \& Binney implicitly probed
by comparing to CO observations.  However, the \hi\ in the neutral ISM
is much less clumpy; it is not clear that the neutral ISM is made up
of discrete clouds, especially over scales of $\gtrsim 50$~pc, the
grid scale we use.  Essentially, applying the Euler equations to the
ISM simply asserts that the ISM has a pressure or sound speed defined
in a coarse-grained sense, over scales greater than the code's
resolution.

Englmaier \& Gerhard (1997) used an SPH code to simulate flow in one
of the model potentials that Athana\-ssoula (1992b) used with the
Eulerian grid code.  For equivalent input parameters, Englmaier \&
Gerhard obtained results very similar to Athana\-ssoula's, which
reassures us that the simulations are not dependent on the
fluid-dynamical algorithm.\footnote{Englmaier \& Gerhard
found that increasing the
sound speed of the gas to 20--25 \kms\ changed the flow pattern.
However, such a large value implies an unreasonably high temperature
for the ISM, and is inconsistent with the value found by Gunn \etal\
(1979).}

A limitation of particle codes is their inability to represent large
density contrasts.  By design, spatially adaptive particle codes
resolve structure well in high density regions, but the finite number
of particles precludes adequate representation of the fluid properties
in very low density regions.  Grid codes, on the other hand, cannot
resolve spatial structure below a few grid cells, but can handle
nearly any density contrast with no increase in overhead, and
represent low and high density regions equally.  In a case such as the
gas in the Milky Way bar, where the geometry and scales of interest
are largely fixed by the stellar potential, spatial adaptivity is
less essential and grid codes are generally
more efficient.  The grid's advantage in density contrast is
especially important since the gas in low density regions will prove
crucial to match the observed emission in the forbidden quadrants of
the \lv diagram, as discussed further in Section
\ref{subsec-invertlv}.

\subsection{Simulation procedure}
\label{subsec-simproc}

We begin each simulation in a quasi-equilibrium state, with the mass
of the bar redistributed in an axisymmetric configuration, the gas on
circular orbits, and a uniform gas surface density of 5 \msurfden.  We
turn on the bar by linear interpolation between the initial
axisymmetric state and its fully barred shape, reaching its final
state in 0.1 Gyr.  The bar growth time is approximately equal to the
orbital period at a radius of 3~kpc.  Different choices for the
growth time and initial density do not particularly affect the
results, save that the final gas density distribution scales 
overall proportionally
to the constant chosen for the initial density. 

We continue the simulation to 0.2 Gyr to allow the gas flow to
``settle'' after the bar has grown, and to 0.3 Gyr to verify that the
flow has stabilized.  The gas response can never reach a completely
steady state, because the gas inside co-rotation continuously loses
energy in shocks and flows toward the center.\footnote{The gas build
up in the center can be significant if the code is run for many
rotation periods, \eg\ several Gyr.  This effect can be lessened by
the use of a ``gas-recycling'' provision in the code.  However, we
found that gas recycling caused long-period oscillations in the flow
with the fine grid used here, probably because it redistributes energy
over the grid (G.\ van Albada, private communication).  The
oscillations do not occur on coarser grids, such as those used by
Athana\-ssoula (1992b), presumably due to the higher numerical
diffusivity.  Since we are not interested in the long-term evolution
of the flow, we avoid this numerical problem by turning gas recycling
off.}  Gas continues to accumulate in the center, but there is very
little change in the gas velocity field from 0.2 to 0.3 Gyr.

We use the gas density and velocity fields at 0.2 Gyr to construct \lv
diagrams as would be seen by an observer in the plane of the
simulation.  The observer is 
placed 8.5 kpc from the Galactic center and in the LSR,
moving with a velocity of $\Theta_0 = 220~\kms$ toward $\ell =
90\deg$, and at a given 
viewing angle -- the angle between the bar major axis and the
Sun-Galactic Center line.  (The effect of a different LSR motion
is discussed below in Section~\ref{sec-lsrmotion}).
The viewing angle is varied to find
the best value, as detailed below in Section~\ref{sec-bestfit}.

For each cell in the simulation grid, we calculate the longitude of
the cell and the angle it subtends, and the radial velocity
of the gas in the cell.  The gas density in the cell and its
distance from the Sun determine the observed brightness.  
The brightness distribution is convolved and sampled in longitude to 
model the angular beamwidth of the telescope and the $0.5\deg$
sampling of the observed positions, and convolved in velocity
to include the effects of the sound speed of the gas ($c_s = 8~\kms$)
and the velocity resolution of the observations (smoothed with
a Gaussian of $\sigma = 5.5~\kms$).

\subsection{Model gravitational potentials}

Our models for the gravitational potential are similar to those used
by Athana\-ssoula (1992a,b).  They have three components: an ellipsoidal
bar, a centrally concentrated bulge, and an extended component to
represent both the disk and halo.

We model the bar as a prolate Ferrers $n=1$ ellipsoid with semimajor
axis $a$ and semiminor axis $b$.  The bar density is given by
\begin{equation}
\rho(x,y,z) = \left\{ \begin{array}{ll} \rho_{0,{\rm bar}} \, (1 -
u^2) & {\rm if}~u^2 < 1, \\ 0 & {\rm if}~u^2 > 1,
\end{array}
\right.
\end{equation}
where
\begin{equation}
u^2 = \frac{x^2}{a^2} + \frac{y^2}{b^2} + \frac{z^2}{b^2}.
\end{equation}

This model for the bar is convenient because its gravitational field
is analytic (Binney \& Tremaine 1987), but it is a crude model for the
real bar (\eg\ Dwek \etal\ 1995).  We compensate for one of its
principal weaknesses by adding a bulge component.  Ferrers bars
are not very centrally concentrated; the bulge component allows 
us to increase the central concentration and to adjust its
strength relative to the bar.
The bulge is a modified Hubble profile sphere
with core radius $r_c$ and density given by
\begin{equation}
\rho(r) = \rho_{0,{\rm bul}} \, \left[1 + \left( \frac{r}{r_c}
\right)^2 \right]^{-3/2}.
\end{equation}
The ``bulge'' component can be viewed as effectively part of the bar;
our treatment of the two as separate analytical components does not
imply that we regard them as distinct, either photometrically or
kinematically.
We use $M_{\rm bul}$ to refer to the bulge mass within 1 kpc of the Galactic
center, since this is most analogous to the central concentration of
the bar; the {\it total\/} mass of a modified Hubble profile sphere
diverges at large radii.

The extended component has the potential
\begin{equation}
\Phi(R) = \Phi_0 \ln\left( 1 + \sqrt{1 + (R/R_c)^2} \right),
\end{equation}
where $R_c$ is scale length.  If all the mass that gives rise to this
potential were to reside in the disk, it would have the surface
density of a Rybicki disk (given by Zang 1976 and derived independently
by Hunter, Ball \& Gottesman 1984):
\begin{equation}
\Sigma(R) = \Sigma_0 \frac{R_c}{\sqrt{R_c^2 + R^2}},
\end{equation}
with $\Phi_0 = 2\pi G\Sigma_0R_c$.  The rotation curve of this
potential becomes asymptotically flat at large radius, making it
suitable for modeling the contribution both of the axisymmetric part
of the stellar disk and of the dark matter halo.

As the simulation is two-dimensional,
it is insensitive to the three-dimensional forms of these
density distributions; any distribution that yielded similar
forces in the plane could be substituted.
Thus, mass can be traded off between the axisymmetric components;
for example, it is unimportant that the density of the Hubble
bulge falls off slowly, since the small 
additional contribution to the rotation curve 
(cf.\ Figure~\ref{fig-axirotcurv}) could be
absorbed into the rotation curve of the disk or halo.

The total potential is specified by seven parameters: a central
density and scale length for each of the bulge and ``disk,'' and a
central density and two axis lengths for the bar.  Our only constraint
is that the rotation curve should be roughly flat outside $R_0$, with
a circular velocity from 200--220 \kms\ at 8.5 kpc.  An eighth
parameter, the Lagrange or corotation radius $R_L$, is required to
fully specify a model; choosing $R_L$ is equivalent to specifying a
pattern speed for the bar.  The gas flow pattern is determined by the
adopted potential, but the \lv diagram further depends on the viewing
angle \philsr\ between the Sun--Galactic center line and the major
axis of the bar.

We varied the parameters by trial and error and examined the \lv
diagrams after each run to learn the effects of changes in bar size,
bar mass, bulge mass, Lagrange radius and so on.  Our goal was to find
a model or models that matched the observations reasonably well,
rather than systematically to explore the parameter space, which is
impractical given the large number of parameters.  We did run some
series to explore the effect of varying a parameter, most notably,
varying the Lagrange radius while holding all other parameters
constant.

In all, we ran 51 models; their parameters are given in Table 1.
The table is sorted by the goodness of fit as
measured by the RMS deviation in velocity between model and data
(discussed further in Section~\ref{sec-bestfit}).  The best fit
viewing angle and the goodness of fit are tabulated in the last two
columns of Table 1.  The models are numbered best
to worst; the number, naturally, does not correspond to the order in
which the models were run, since we improved the models by learning
from past results -- Model 1 was actually the 46th model run.



\section{Comparison to observations}
\label{sec-compar}

We compared the outer envelope -- the extreme-velocity contour -- of
the synthesized \lv diagrams to that of the data.  The observed EVC
used is the contour of 0.125~\degk~degrees of antenna temperature
summed over $b$, or 1.25~\lvunit\ of atomic gas (H and He), using the
calibration given by Liszt \& Burton (1980).  The data points used are
shown by the filled circles in Figure \ref{fig-lisztdata}.  As
discussed in Section~\ref{sec-lvdiag}, those portions of the EVC that
show signs of contamination from foreground disk emission are excluded
from comparisons to models.

\subsection{The EVC contour level}  
\label{sec-evclevel}

The position of the observed extreme-velocity contour is determined 
by the actual terminal velocity envelope of the gas, extended
by the velocity broadening due to the gas sound speed and
the instrumental resolution.  Since the flux level at which the EVC can be
observed is also limited by the noise in the observations,
the EVC is not an intrinsic property of the Galaxy, but also depends
on the observational parameters.  
In the tangent-point method, the observed EVC must be corrected
to yield the terminal velocity envelope.  In practice, it is
conventional to assume that (1) the difference between the
terminal velocity $v_t(\el)$ and the EVC
is some constant $\Delta V$, 
and (2) $\Delta V$ can be determined by observations near 
$\ell \simeq \pm 90\deg$,
where the actual terminal velocity is expected to be zero
(cf.\ Gunn \etal\ 1979).

As the data we are using do not cover $\ell \simeq \pm 90\deg$,
we cannot make use of this method to derive $\Delta V$.
In order to compare the observations and simulations, we have
constructed simulated \lv diagrams which take into account
the velocity dispersion of the gas and the instrumental resolution.
But the absolute level at which to place the EVC in the 
simulated \lv diagram is not constrained, since we do not 
know $\Delta V$ for the observations.  Fortunately, both simulated and
observed \lv diagrams have fairly sharp edges, in the sense
that the flux falls off rapidly with increasing $|V|$ --
see Figures \ref{fig-lisztdata} and \ref{fig-bestlv}.  
The lowest contours simply trace the falloff profile of the 
velocity dispersion and instrumental resolution.
We compared simulated \lv diagrams to that observed, 
examining the fall off at the edges of the distribution,
to set the level for the EVC in the simulated \lv diagram.
Placing the EVC at $\sim 1.7$ \lvunit of total simulation gas 
produced a reasonably good match
but EVC levels of 1.25 -- 2.5~\lvunit\ were almost equally acceptable.   

Because the \lv diagrams do have sharp edges, changing the 
flux level of the comparison EVC, even by a factor of 2, does not 
have a strong effect.  We ran comparisons of the entire series of models
at EVC contour levels from 0.625 -- 5.0~\lvunit\
and verified that the small changes
caused by different choices for the EVC level
produce only minor changes in the
rank ordering of models, and do not affect our conclusions.

We note that comparing the simulation EVC at 1.7~\lvunit\
of total gas to the observed \hi\ EVC at 1.25~\lvunit\ of atomic gas
could be interpreted to mean that the gas is 75\% atomic; levels
of 1.25 -- 2.5~\lvunit\ would imply atomic fractions of 100\% -- 50\%. 
However, the comparison is not reliable for this purpose.
The edges of the EVC largely represent gas in
low density regions, and the molecular fraction is undoubtedly
higher in high density regions -- CO emission does not generally
extend to the velocities of the \hi\ EVC (Dame \etal 1987).
Additionally, the inferred fraction would be changed if the value
of the initial gas surface density used in the simulations were
changed.  It is, however, comforting that the inferred atomic fraction
is close to but less than 1.  (The actual atomic mass fraction
in the inner Galaxy is perhaps 50\%; cf.\ Bronfman \etal\ 1988; 
Bloemen \etal\ 1986.)

\subsection{Best fit and viewing angle}
\label{sec-bestfit}

To rank the models by the quality of their fit to the data, we compute the
root-mean-square deviation in velocity between the location of the
simulated EVC and the observed data points.  The RMS velocity
deviation is not an ``error'' in a statistical sense; it serves as a
figure-of-merit for ranking the models.  The RMS places a relatively high
weight on large deviations, which penalizes gross differences between
model and data more than large numbers of small differences.

For a given model, the position of the observer with respect to the
bar must be specified to construct an \lv diagram.  We define the
``viewing angle'' \philsr\ to be the angle between the bar major axis
and the Galactic Center--to--Sun line, so that 0\deg\ is an end-on
bar, 90\deg\ is side-on, and values between 0\deg\ and 90\deg\ put the
near end of the bar in the first Galactic quadrant ($0\deg < \ell <
90\deg$).  We determined the best-fit viewing angle for each model
iteratively by synthesizing \lv diagrams and computing the RMS
deviation at viewing angle intervals of 10\deg, 4\deg, and 1\deg,
successively, narrowing the search interval at each step.  The
best-fit viewing angle for each model and the corresponding RMS
velocity deviation are tabulated in Table 1,
sorted by the goodness of fit.  The viewing angle given is for the
best fit between 0\deg\ and 90\deg; these are the realistic models
since many lines of evidence place the near end of the bar in this
quadrant.  For the few models that have a better fit outside this
quadrant, that result is given in the table footnotes.


\section{Results: I. The best model}
\label{sec-results1}

\subsection{Properties of the model}
\label{sec-propmodel}

\begin{figure*}[ht]

\psfig{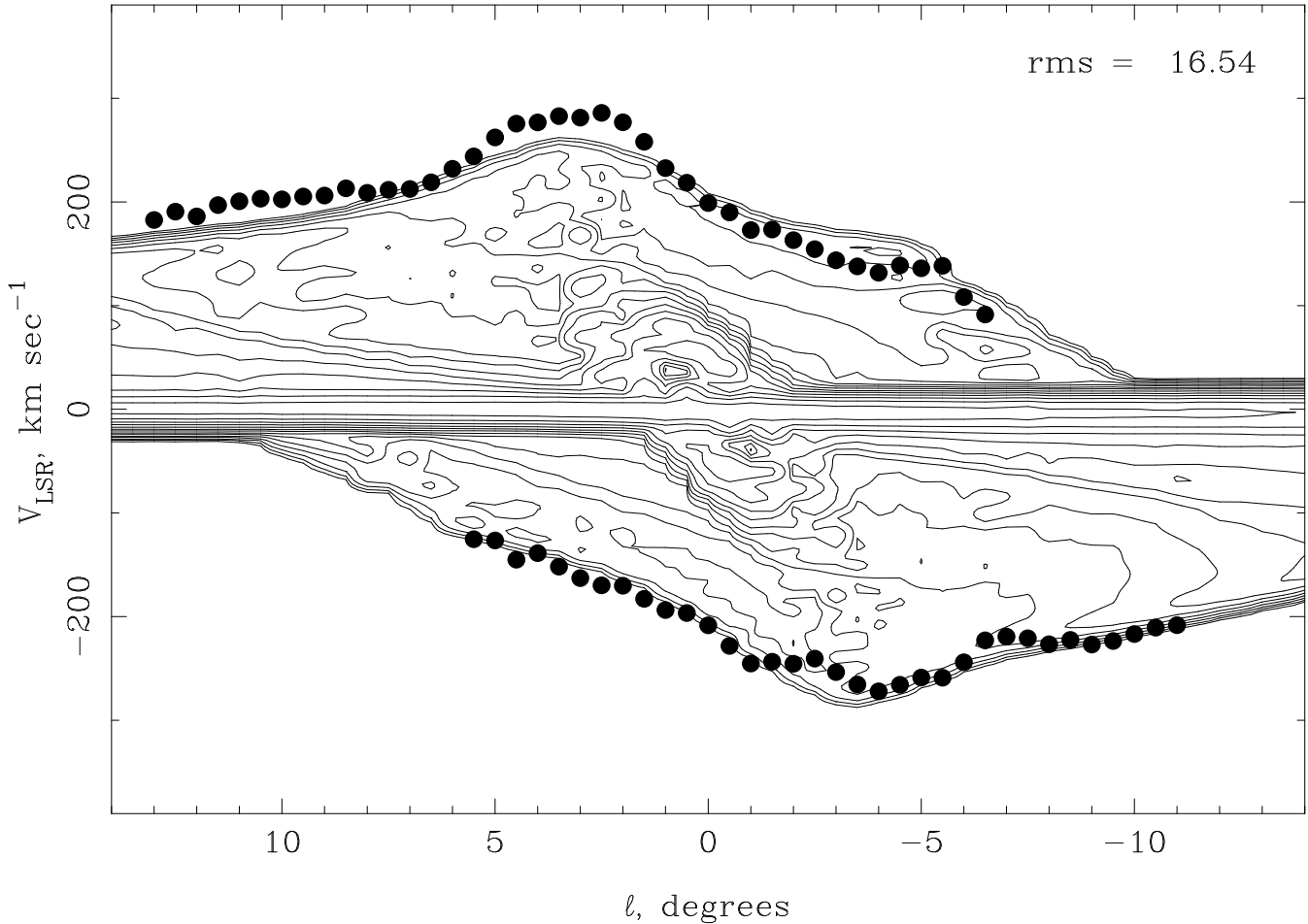}

\caption{The \lv diagram for the best-fitting model, Model 1, as
viewed at 34\deg.  The lowest contour is 1.7 \lvunit\ of total gas and
the contours increase geometrically by a factor of 2.  The filled
circles are the observed data points.  The RMS velocity deviation
between model and data is 16.54~\kms.
\label{fig-bestlv}
}

\end{figure*}

Our primary result is that we have found a model which 
reproduces the outer contour of the \lv diagram fairly
well.  This model is model 1 in Table 1;
a number of the models that are runners-up are closely
related to it.  Model 1 has a bar
with semimajor axis 3.6 kpc and Lagrangian radius 5.0 kpc,
corresponding to a pattern speed of 41.9 \kmskpc.  The best-fit \lv
diagram is shown
in Figure \ref{fig-bestlv} and the RMS velocity deviation is 16.54~\kms.
The minimum in RMS deviation is well localized at a viewing angle of
34\deg\ to the bar major axis, although changes of a
few degrees ($< 5\deg$) are possible without greatly worsening the
fit.  The localization in RMS deviation is similar
for all of the better models.  The effects of changes
in the viewing angle are discussed further in Section
\ref{sec-results2}.  

\placefigure{fig-bestlv}

\begin{figure*}[ht]
\psfig{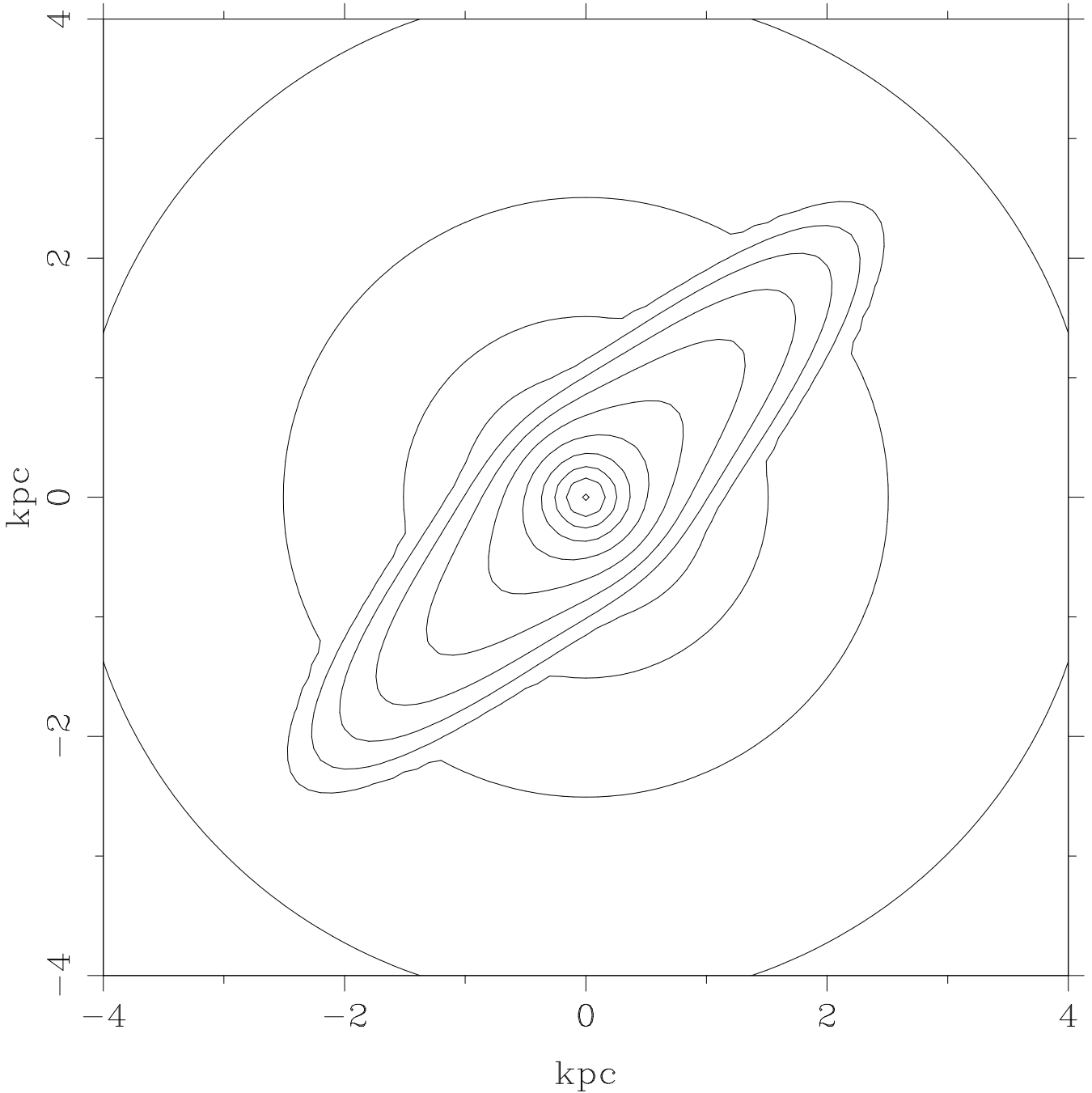}

\caption{The face-on surface density of the mass distribution for
Model 1, as projected along the $z$-axis, and shown for the innermost
8 kpc by 8 kpc of the Galaxy.  The lowest contour is at
400~\msurfden, and the contours increase geometrically by
a factor of $\sqrt{2}$.
\label{fig-surfmass}
}

\end{figure*}

Figure \ref{fig-surfmass} shows the surface density
distribution of the combined bar, bulge, and disk+halo components,
as projected along the $z$ axis of the Galaxy.
The figure shows the central 8 kpc by 8 kpc region of the model, in
what is essentially a face-on view -- although the contours are in
surface density of mass, not light.  This plot demonstrates
the influence of the bulge component, which makes the central
concentration of the bar much higher than that of a Ferrers bar in
isolation.  It is also clear that the full surface density
distribution is less elongated than the bar component alone, with an
axis ratio of about 3:1 as compared to the bar component's 
axis ratio of 4:1.

\placefigure{fig-surfmass}

When the mass density is integrated over $-100 < z < 100$~pc,
the lower estimate for the thickness of the gas layer,
the resulting distribution is similar to that of Figure
\ref{fig-surfmass} but with mass surface density lower by
a factor of about four.
Within this range of $z$, at the bar end the mass surface density
is 140~\msurfden; in the central region at $R<0.5$~kpc the 
mean mass surface density
is 2900~\msurfden.  A comparison of these mass surface
densities with the gas surface density suggests our neglect
of gas self-gravity is justified.

\begin{figure*}[htb]

\psfig{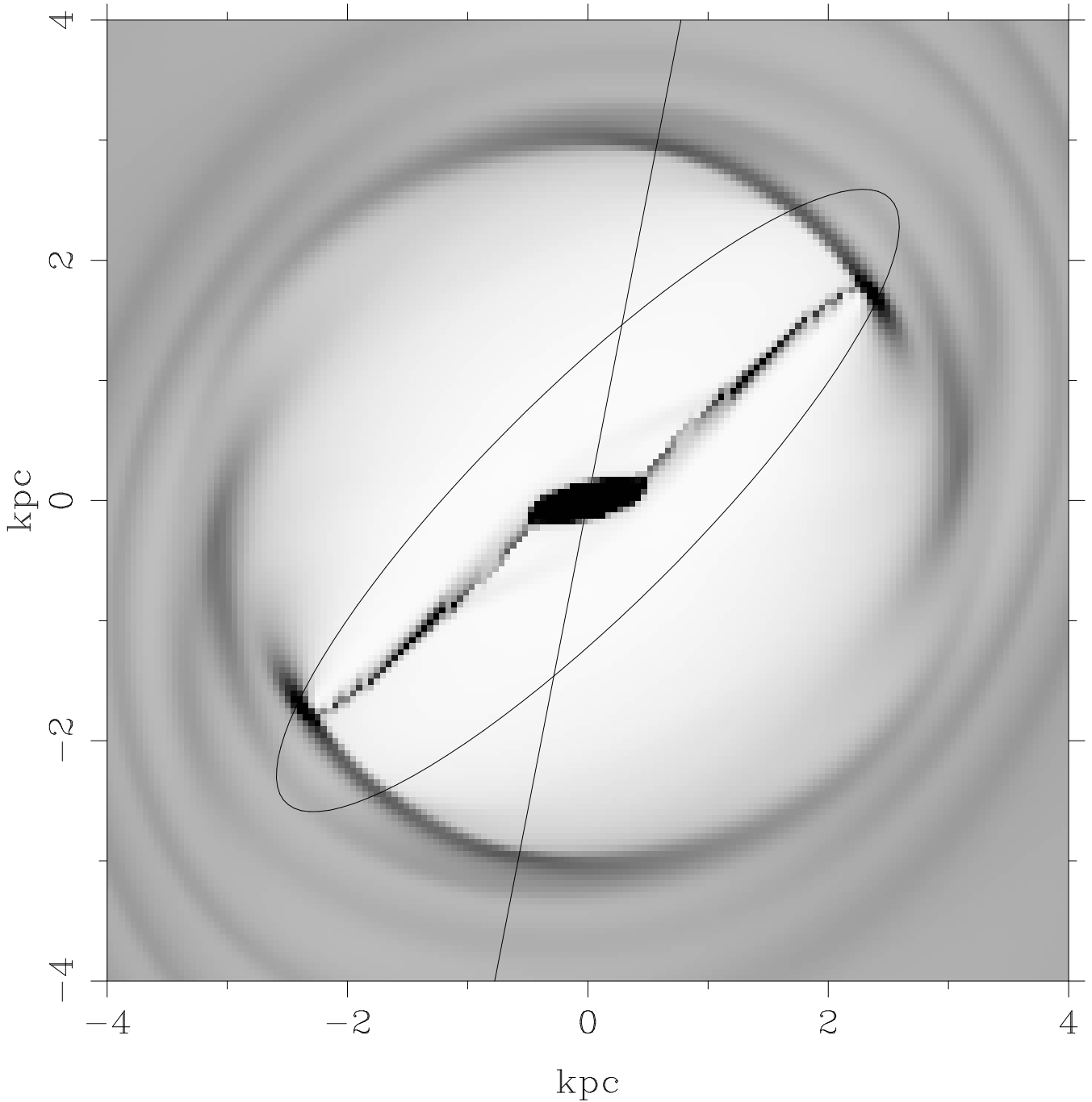}


\caption{The face-on gas density in the innermost 8 kpc by 8 kpc
of Model 1, as viewed from above the plane of the simulation. 
The grayscale runs from 0 to 15 \msurfden\ of total gas, with
dark representing higher density.  The bar
is at a 45\deg\ angle to the grid, running from lower left to upper
right.  The ellipse represents the points at which the bar density
drops to zero, with semimajor axis 3.6 kpc.  The sense of the bar
rotation is clockwise.  The inclined line from top to bottom of the
figure represents the Sun--Galactic Center line for the best-fit
viewing angle of 34\deg, with the Sun above the top of the figure.
\label{fig-gasdensell}
}

\end{figure*}

\placefigure{fig-gasdensell}

The gas density in the innermost 8 kpc square of this model is shown
in Figure \ref{fig-gasdensell}, also in a face-on view.  
The long, straight high density
features in the bar are shocks, with transverse velocity jumps of
$\sim 200~\kms$, extending out to 2.9 kpc from the galactic center.
They are parallel but offset from each other; near the center the
straight shocks join onto an oval or nuclear ring
of high density gas that is also
the location of shocks.  The semi-major axis of this oval is
0.5 kpc.  The gas surface density in the shocks is 5--20~\msurfden;
within $R<0.5$~kpc the mean gas surface density is 130~\msurfden.
The straight, offset shocks and the inner oval
are characteristic of gas
flow in strongly barred potentials with an inner Lindblad resonance
(Athana\-ssoula 1992b).

Dust lanes with morphologies similar to the high density gas
in Figure \ref{fig-gasdensell} are
observed in many barred galaxies.  The dust lanes are presumably
caused by the high gas density at the shock (Prendergast 1962,
unpublished; 
see also van Albada \& Sanders 1982; Prendergast 1983; Athanassoula 1992b).
Spectroscopy of barred galaxies shows sharp
velocity jumps at the location of the dust lane (\eg\ Pence \&
Blackman 1984; Lindblad \etal\ 1996; Regan, Vogel \& Teuben 1997; Weiner
\etal\ 1999).

Figure \ref{fig-gasdensell} also shows that there is little gas in the
lens region of the galaxy, inside 3~kpc; barred galaxies often show a
central hole in the gas distribution swept clear by the angular 
momentum transport of the bar (\eg\ NGC 1300, England 1989; 
NGC 1398, Moore \& Gottesman 1995; and NGC 4123, Weiner \etal\ 1999).
Outside 3~kpc,
the gaseous disk is relatively quiescent; the bar does not drive
a large response in the outer disk.  The disk does
not exhibit spiral patterns outside the bar radius; spirals
in the outer disk could be driven by spirals in the stellar disk
and/or the self-gravity of the gas, which we have neglected in
order to concentrate on the inner Galaxy.  

\begin{figure*}[ht]

\psfig{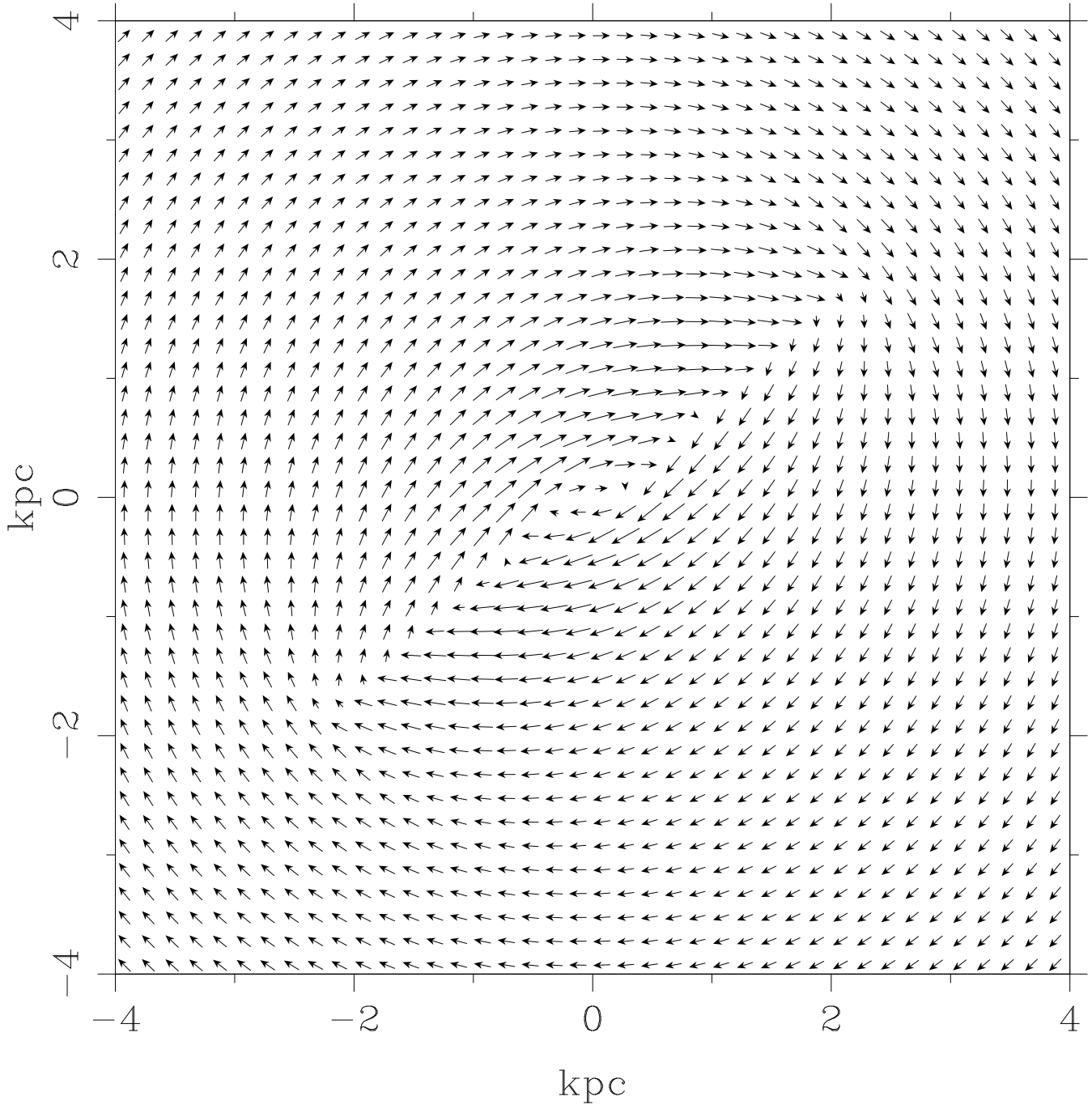}

\caption{The gas velocity field in the innermost 8 kpc by 8 kpc of
Model 1.  The bar is at a 45\deg\ angle to the grid as before.
For clarity, only every fourth cell in the grid is plotted.
\label{fig-streamlines}
}

\end{figure*}

\placefigure{fig-streamlines}

The gas velocity field as seen in a non-rotating frame, in the inner
$8 \hbox{ kpc} \times 8$ kpc region, is shown in Figure
\ref{fig-streamlines}.  For clarity, we have plotted only every fourth cell.
The velocity changes abruptly at the shocks along the density peaks.
Essentially, gas in the bar moves up to the shock at relatively high
velocity and hits the shock, dissipating energy.  The post-shock gas
then streams back down the bar, gaining velocity quickly as it moves
away from the shock and falls down the potential well.  

Gas streamlines in the bar are elongated along the bar, in the manner
of the $x_1$ family of stellar orbits in bars, but are clearly not
symmetric about the major axis of the bar, unlike the $x_1$ orbits.
The shocks 
are located along the leading edge of the streamlines and are
approximately parallel to the bar; the major axis of the 
elongated streamlines is rotated approximately $5\deg$ 
ahead (toward the leading side) of the bar major axis.
This angle, which we will refer to as the ``lead angle,''
is closely related to the pattern speed of the bar, to
be discussed further in Section~\ref{sec-patspeed}.

Near the center of the bar, the major axes of the streamlines change,
so that the streamlines are elongated across the bar more than along
it, similar to the $x_2$ family of stellar orbits present in bars with
inner Lindblad resonances (Athana\-ssoula 1992a,b).  The central oval of
high gas density corresponds to this family of streamlines.
Again, the streamlines are rotated by an oblique angle with respect 
to the bar, unlike the $x_2$ stellar orbits, which are perpendicular
to the bar major axis.

\subsection{Inverting the projection into \lv space}
\label{subsec-invertlv}

\begin{figure*}[ht]

\psfig{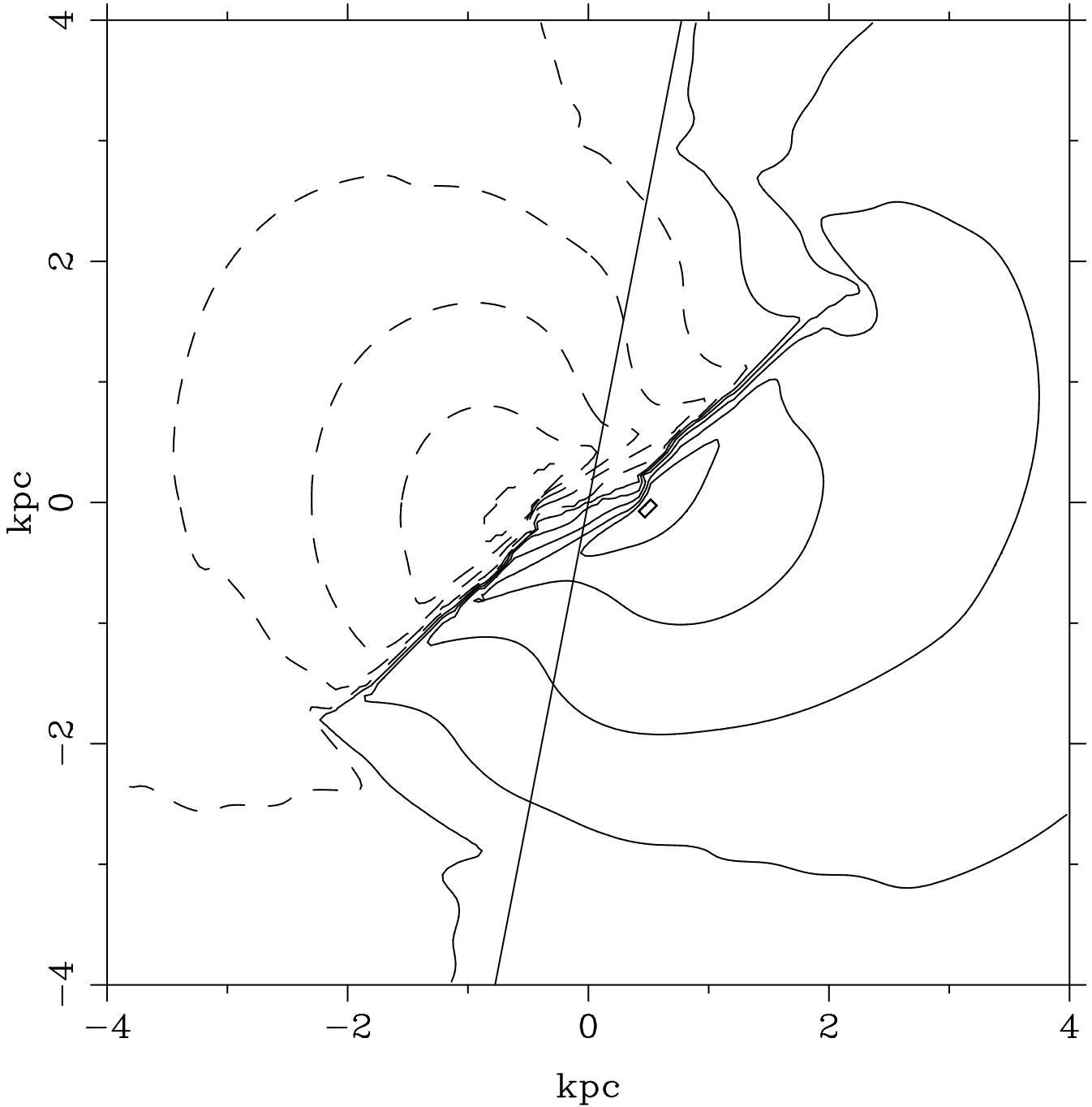}

\caption{A contour plot over the inner $8 \hbox{ kpc } \times 8$ kpc region
of the radial velocity at which each grid cell in Model 1 is observed when
the viewing angle is 34\deg.  The contours are from --250 \kms\ to 250
\kms\ at intervals of 50 \kms; contours below zero are dashed.  The bar is
at a 45\deg\ angle to the grid, as before.  The inclined line from top
to bottom is the Sun--Galactic Center line ($\ell = 0\deg$), with the
Sun above the figure.
\label{fig-velcontour}
}

\end{figure*}

The plot of the gas streamlines offers some understanding of the
features in the \lv diagram, but the effect of projection into \lv space
is much clearer in Figure \ref{fig-velcontour}.  This figure plots
the radial velocity observed in Model 1 as a function of position on the
grid, i.e.\ over the plane of the Galaxy, showing
the radial velocity {\it before\/} it is projected into
the \lv diagram.

\placefigure{fig-velcontour}

The gas at forbidden velocities moves toward the Sun at $\ell >
0\deg$, the side where most of the gas is moving away, and vice versa
at $\ell < 0\deg$.  It is clear from Figure \ref{fig-streamlines} and
Figure \ref{fig-velcontour} that forbidden velocities belong to
low-density gas approaching the shocks.  The preshock region
with forbidden velocities extends all the way
out to the shock tip at 2.9 kpc.  However, the magnitude of the
forbidden radial velocities in the preshock region falls below
100~\kms\ at about 1.5 kpc from the Galactic center, which corresponds
roughly to $\ell = \pm 6\deg$ for the viewing angle of 34\deg.  Past
this point, emission at forbidden velocities is obscured in the 
Milky Way by the band of emission from foreground gas.

Identifying the emission in the forbidden quadrants with the
low-density preshock gas may explain why the forbidden emission is
much more extensive in \hi\ than in CO, while the peaks in the \hi\
\lv diagram,
which come from higher density regions, are present in the CO 
\lv diagram.  The identification of forbidden velocities with
the preshock gas also
illuminates some of the difficulty Jenkins \& Binney (1994) had
matching their sticky-particle models to the data.  The \lv diagrams
they presented have very little emission in the forbidden quadrants.
However, their maps of gas density in the plane of the simulation show
that the apparent lack of emission is because there are very few
particles in the preshock regions at any given time, as discussed in
Section \ref{sec-simul}.

The high and narrow peaks in the EVC at
$\ell \sim +3\deg \hbox{ and } -4\deg$,
or $\sim 0.6$~kpc projected distance from the Galactic center, have no
counterparts in the equivalent axisymmetric model 
(Figure~\ref{fig-axisymlv}).  The
origin of these peaks can also be understood from Figure
\ref{fig-streamlines}; the elongation of the orbits caused by the
strong ellipticity of the gravitational potential results in high gas
velocities roughly parallel to the bar major axis.  The observed high
radial velocities arise from the gas on elongated orbits
just as it passes the oval
of high-density gas (Figure \ref{fig-velcontour}).  The EVC
declines rapidly beyond the peak because the gas at larger radii does
not fall as deeply into the bar's potential well, and is on less
elongated orbits.  

Many authors (e.g.\ Gunn \etal\ 1979; Gerhard \& Vietri 1986;
Liszt 1992; Burton \& Liszt 1993) have noted that the peaks in
the EVC and the rapid decline imply an unusual rotation curve if the gas is
assumed to move on circular orbits; the inferred rotation curve also shows
a sharp rise and rapid decline.  These features are more naturally
explained by gas flow in a triaxial potential (\eg\ Gerhard \& Vietri 1986,
Burton \& Liszt 1993).  
Simulations such as model 1 show that not only the EVC peaks, but also
the forbidden emission, are accounted for by gas flows in a strong bar.
As Burton \& Liszt emphasize,
comparisons with a derived rotation curve instead of the full \lv diagram
both embody incorrect assumptions about the
inner Galaxy and discard valuable data from
the forbidden quadrants of the \lv diagram.

Figure \ref{fig-velcontour} can also be used to
determine the location within the plane of the Galaxy of
a feature in the \lv diagram, or an object whose
longitude and radial velocity are known but whose distance
is uncertain.  For example, the 3-kpc expanding arm
goes approximately through the points 
$(\el,V)$ = ($-10\deg$, $-100~\kms$), ($-5\deg$, $-75~\kms$),
(0\deg, $-50~\kms$), (+2.5\deg, $-35~\kms$) (Liszt \& Burton 1980).  
Locating these points on Figure 
\ref{fig-velcontour} shows that they lie approximately on a arc 
centered on the Galactic center and of $\sim 2.5$~kpc radius, 
suggesting that the 3-kpc arm could be a spiral arm at about that radius 
with a small pitch angle, and that its motion is consistent
with the overall Galactic velocity field, removing the
need for large anomalous expansion velocities.  In fact, an arm
at approximately the right position is visible in
Figure~\ref{fig-gasdensell}.

We note that even though the simulation is bisymmetric, 
the synthesised \lv diagram
has some asymmetry because one end of the bar is closer to the Sun
than the other.  The observed \lv diagram is somewhat more asymmetric
than the model, however.  We cannot rule out the possibility that the
observed asymmetry is due to actual asymmetries in the gas
distribution or the shape of the Galaxy.  However, the asymmetries in
\hi\ are considerably smaller than those in the CO \lv diagram (Dame
\etal\ 1987; Bally \etal\ 1988).  

The most obvious deviation of this model from the observations
is that it is not as strongly
peaked at positive \el\ as the data, although this is essentially due
to the asymmetry in the peaks of the data, since the model compromises
by slightly overestimating the peak at negative \el.  The model
also produces a strong diagonal feature in the interior of the 
\lv diagram from about (+3\deg, +100~\kms) to (--3\deg, --100~\kms)
which is not present in Figure~\ref{fig-lisztdata}.  
This feature is caused by the nuclear
ring of high density gas, which is not seen in the observations both because
the gas is probably in the $H_2$ phase (cf.\ Rubin,
Kenney \& Young 1997) and because those parts of this feature with
$|V|<100\;$km~s$^{-1}$ are obscured by the ``main
maximum'' of foreground from disk gas.

The gas density
and velocity fields in our model 1 are consistent with those observed in
external barred galaxies.  In particular, the straight shock regions
with high gas densities can be identified with the straight dust lanes
along the bar seen in many barred galaxies, which are generally
thought to be the locations of shocks (Prendergast 1962, unpublished;
see also \eg\ Prendergast 1983; van Albada \& Sanders 1982;
Athana\-ssoula 1992b).

Model 1 provides the best fit among our models, but it is by no
means a unique solution to the problem of reproducing the
\lv diagram.  A slightly different choice of parameters
could conceivably do better, and it is almost certain
that some potential with components other than the particular analytic
forms we used could improve on Model 1.
However, the Galaxy is likely to resemble Model 1
in certain major respects, such as viewing angle, bar size, possession
of an ILR, and high pattern speed.  These conclusions are partly drawn
from our experience with other, less well-fitting models, which we now
discuss.

\section{Results: II. Other models}
\label{sec-results2}

In this section we describe other models to illustrate the
influence of variations in some of the major parameters.  This
exercise allows us to infer the properties that a successful
model is likely to possess in order to reproduce the observations.

A natural question to ask is whether 
the adverse consequences of changing one parameter can
be compensated for by changes to other parameters.
In general, the effects of the parameters are sufficiently 
interlinked that attempting to compensate by making one change 
has other unintended consequences.  Given the number of parameters,
it is impractical to test for all possible compensatory changes,
but we do not believe that large variations in important parameters can
be compensated away.
Although Model 1 was one of the last models
to be run, it is a close variant of Model 3, which we had
tried much earlier (our 24th run);
we tried a number of variations to improve Model 3 before
actually succeeding.  Our
experience makes it seem unlikely that some other radically
different model could fit equally well or better, but we
cannot rule out the possibility.

\subsection{Changes in viewing angle}
\label{sec-viewangle}

\begin{figure*}[htb]

\psfig{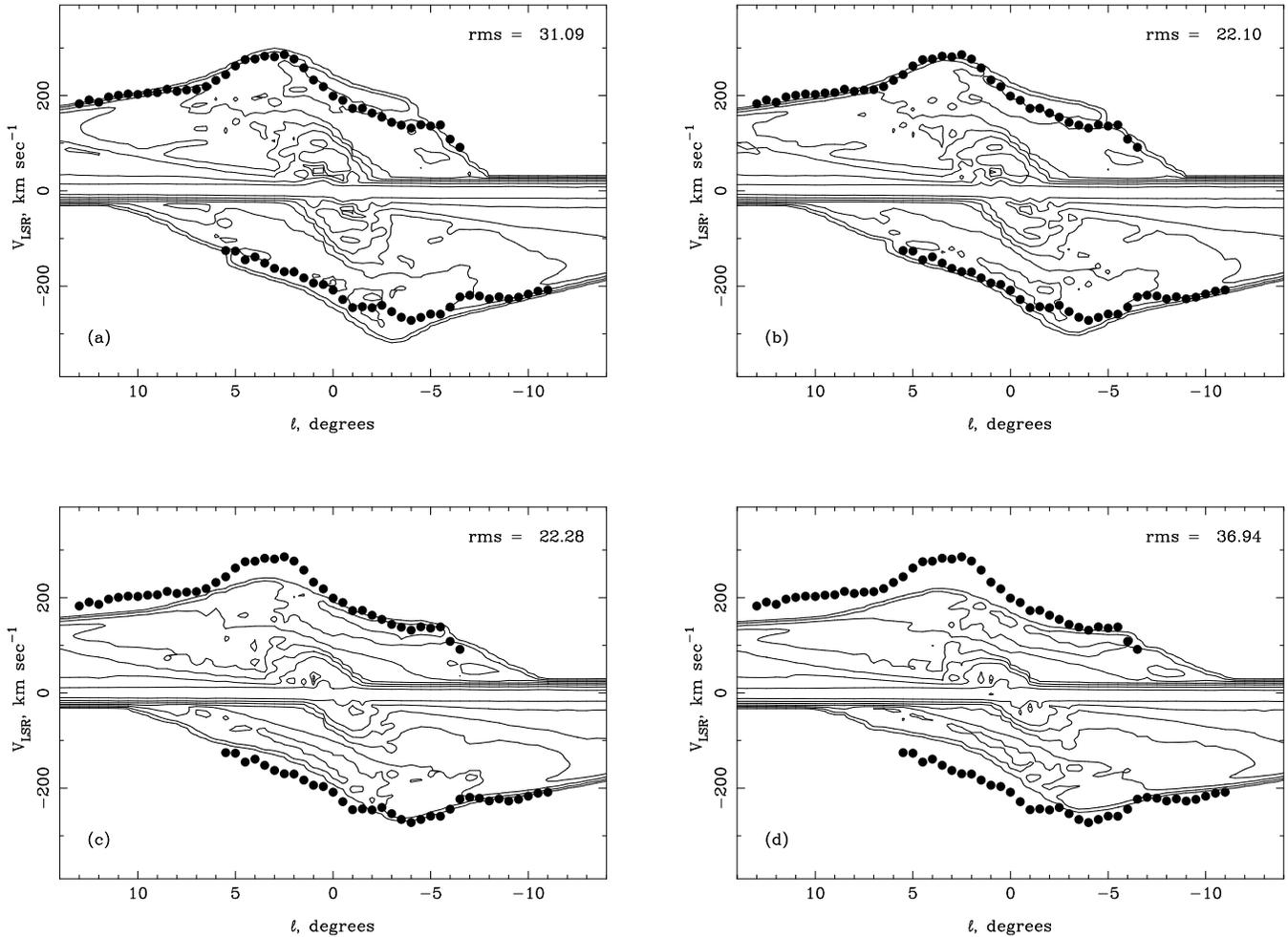}

\caption{Longitude-velocity diagrams for Model 1 when it is viewed at
angles other than the optimal viewing angle of $\philsr = 34\deg$.
The lowest contour is 1.7 \lvunit\ of total gas and the contours
increase geometrically by a factor of 4.
\label{fig-viewangle}}
(a) $\philsr = 24\deg$, more end-on; RMS velocity deviation = 31.09~\kms.\\
(b) $\philsr = 29\deg$,  RMS = 22.10~\kms.\\
(c) $\philsr = 39\deg$, more side-on; RMS = 22.28~\kms.\\
(d) $\philsr = 44\deg$, RMS = 36.94~\kms.

\end{figure*}

Changing the angle from which Model 1 is viewed is not properly a
different model for the potential, but can drastically change the
resulting \lv diagram.  Figure \ref{fig-viewangle} illustrates the
systematic changes that occur when Model 1 is viewed at angles
$-10$\deg, $-5$\deg, $+5$\deg, and $+10$\deg\ from the optimum
value of $\philsr = 34\deg$.

Viewing a model more nearly end-on than optimum, as in Figure
\ref{fig-viewangle}(a) and (b), produces both higher peaks in the EVC
and steeper declines from the peaks.  It also reduces the extent of
the gas in the forbidden quadrants, relative to the
height of the peaks.  Once again, reference to Figure
\ref{fig-streamlines} reveals the reasons for these changes.

\placefigure{fig-viewangle}

In a more end-on view of the bar, the elongated orbits that produce
the velocity peaks are projected more onto the line of sight of the
observer, making the peaks higher.  Counter to what might be
expected, the peaks do not move significantly closer together in a
more end-on view because the streamlines in this part of the flow are 
curved, and the region contributing to the peaks rotates
somewhat. 
The curve of the streamlines is caused by the presence of an ILR,
because the $x_2$ orbit
family forces the elongated streamlines in the inner region of the bar
away from the center.  The more end-on view also means that the region
with highly to moderately elongated orbits subtends a smaller angle,
and so the fall-off with increasing $|\el|$ is more rapid.

The relative deficiency of gas in the forbidden quadrants occurs
because the shocks, and the preshock regions responsible for the
forbidden emission, subtend a smaller angle when the bar is viewed
more end-on.  In the more end-on view, the projected components of the
velocities of the preshock gas are larger, which compensates somewhat,
but the slope of the decline in the EVC from the peaks into the
forbidden quadrants is steeper.  Clearly, the peaks could be lowered
by reducing the central density of the model, but the more end-on view
would then yield too little emission in the forbidden quadrants.

The effects of a more side-on view of the bar, as seen in Figure
\ref{fig-viewangle}(c) and (d), are essentially exactly the opposite.
The velocity peaks drop and their slope is gentler.  The extent of the
gas in the forbidden quadrants increases, but the lower projected
velocities give a gentler slope to the EVC.

Gross variations in viewing angle, to the point where, for example, a
model is viewed fully side-on at $\philsr \sim 90\deg$, can produce
\lv diagrams that deviate somewhat from these rules of thumb. For
example, some models such as numbers 6, 9, 12, and 18
can produce high velocity peaks at side-on viewing
angles because the innermost streamlines derived from $x_2$ orbits
(approximately perpendicular to the bar) are viewed end-on.
These models are of little practical interest, since a number of other lines
of evidence rule out such large viewing angles -- for example, a
grossly side-on view cannot produce the magnitude offset between bulge
stars at positive and negative longitudes, as shown by Stanek \etal\
(1997).  (We note that models 6, 9, 12, and 18 are all slow bars
in which $R_L \geq 2.4a$; see below.)

\subsection{Motion of the LSR}
\label{sec-lsrmotion}

The \lv diagrams were constructed by assuming that the
LSR is moving with a circular (tangential) velocity
$\Theta_0 = 220$~\kms\ relative to the Galactic Center, with no 
radial motion.  We tested the effect
of assuming a different velocity of the LSR relative to the
Galactic Center.  A radial motion of --5 to +10~\kms,
positive outward, can be accommodated; values 
outside this range significantly worsen the models' fit
to the data.  The best values of the radial motion are between
0 and +5~\kms.  The fits are not sensitive to
reasonable variations of the circular speed,
since the data are near $\ell  = 0\deg$; values
of $\Theta_0$ from 160 to 240 \kms\ were tested
and yielded acceptable fits.  Varying the LSR motion
has a minimal effect on the relative ranking of the models.

The non-circular motion predicted by the models for gas 
at the solar position is small.  For Model 1, the 
gas at the solar position has a tangential velocity
of 211~\kms, and a radial motion of --0.7~\kms\
(inward).  Model 1 has an outer Lindblad resonance (OLR)
near the solar position, but the gas is on an essentially
circular orbit.  The OLR could have observable effects on the
kinematics of stars in the solar neighborhood, in either 
mean velocity or dispersion.  The nature
of the effects is not simple to predict
(cf.\ Kalnajs 1992, Kuijken \& Tremaine 1992, Weinberg 1994);
moreover Dehnen's (1998) analysis of Hipparcos data shows that the 
velocity structure of nearby stars is quite complicated.

\subsection{Varying the pattern speed}
\label{sec-patspeed}

\begin{figure*}[htb]

\psfig{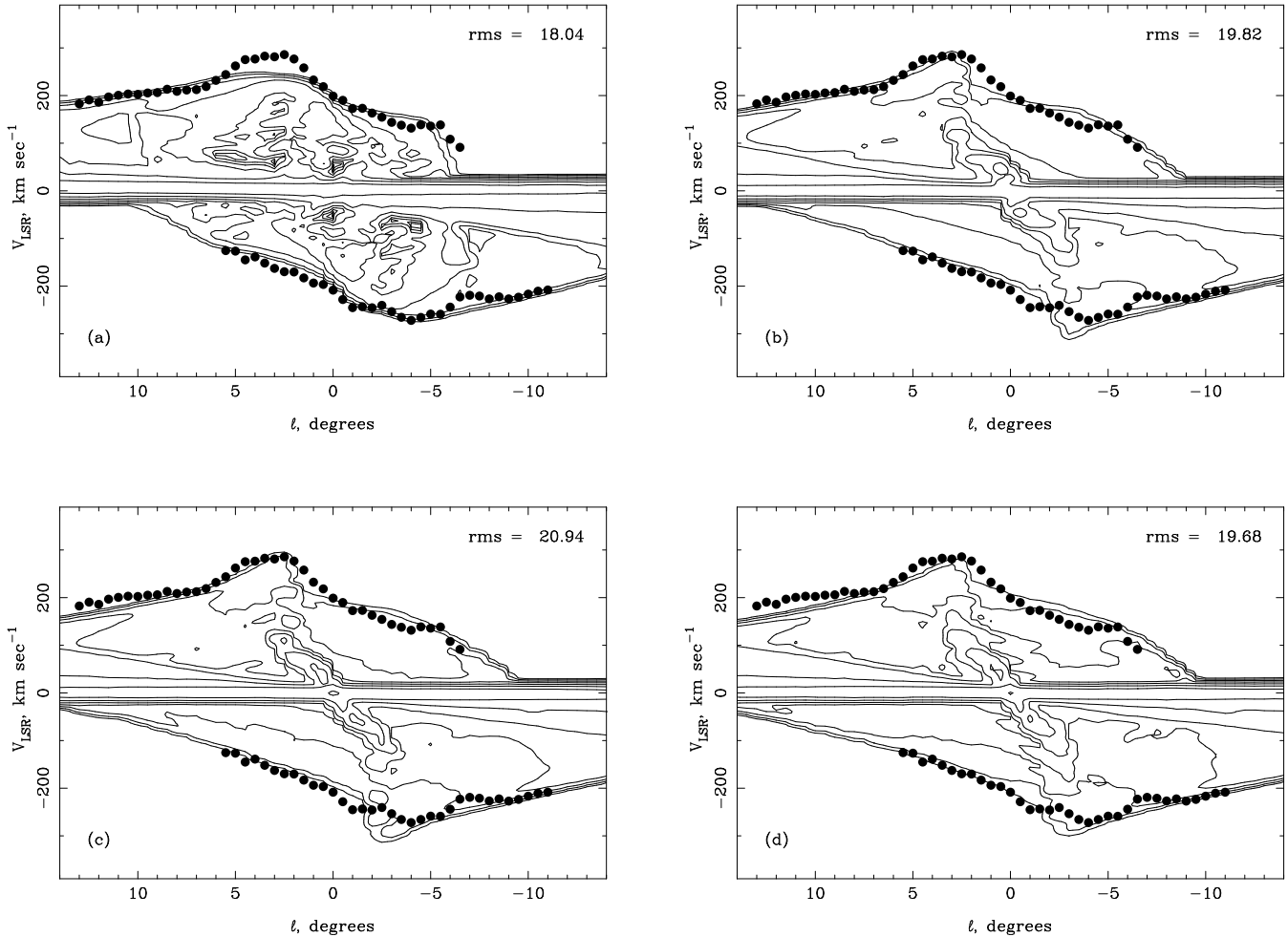}

\caption{Longitude-velocity diagrams for a sequence of
models with the same parameters as Model 1 except for
the Lagrange radius.  Model 1 (Figure \ref{fig-bestlv})
is between Models 2 (panel a) and 5 (panel b) in the sequence
of increasing $R_L$.
The lowest contour is 1.7 \lvunit\ of total gas and the contours
increase geometrically by a factor of 4.
\label{fig-radlaglv}}
(a) Model 2, $R_L$ = 4.0 kpc,
bar viewing angle $31\deg$. \\
(b) Model 5, $R_L$ = 6.0 kpc,
bar viewing angle $20\deg$. \\
(c) Model 8, $R_L$ = 7.0 kpc,
bar viewing angle $15\deg$. \\
(d) Model 4, $R_L$ = 8.0 kpc,
bar viewing angle $8\deg$. 

\end{figure*}

\begin{figure*}[htb]

\psfig{figure=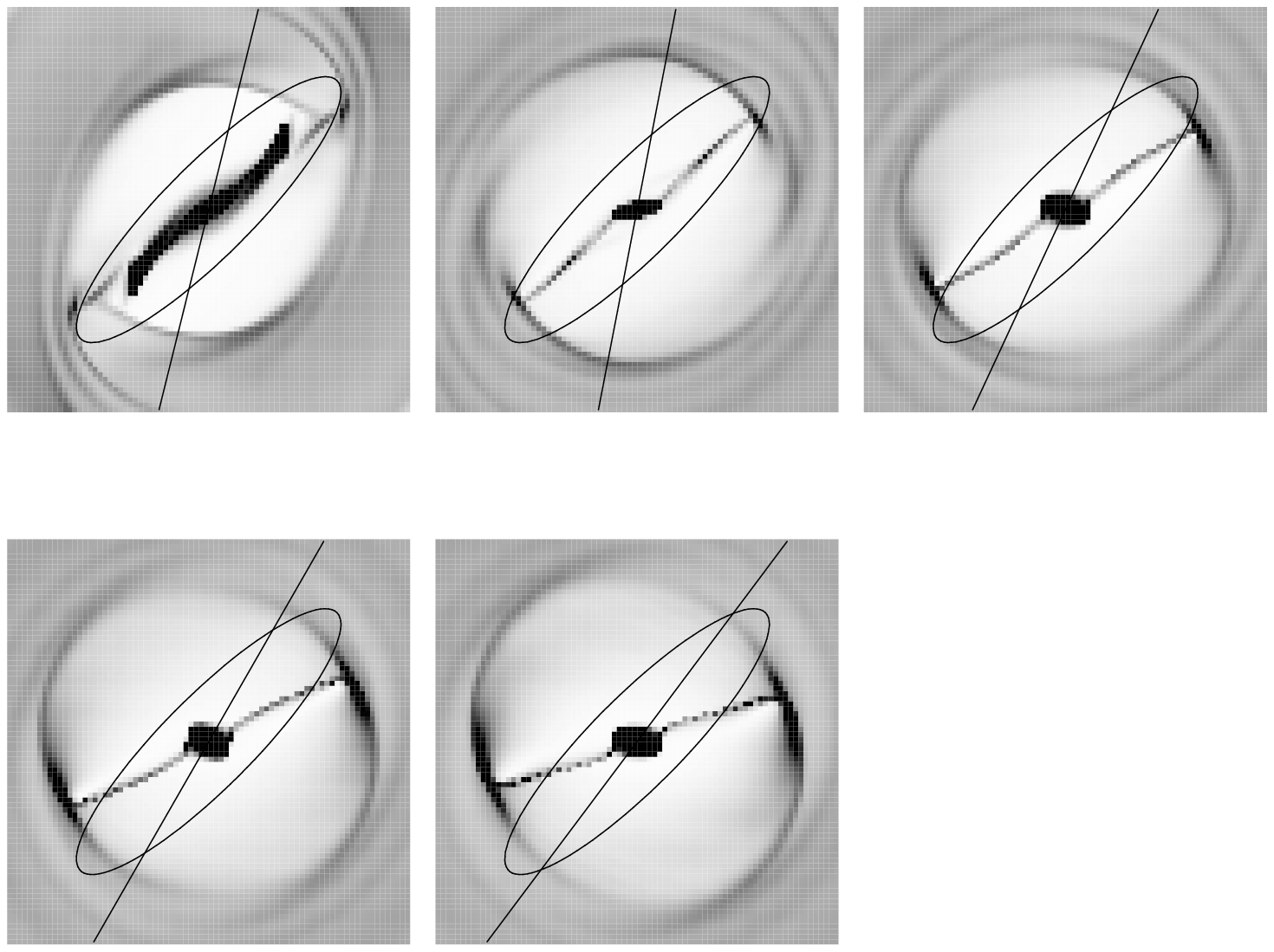,width=7.0truein}


\caption{The gas density in the inner 8 by 8 kpc of the grid
for a sequence of five models with different Lagrange radii
and all other parameters fixed.  The bar is at 45\deg\
to the grid in each model; the ellipse indicates the outer edge
of the bar, at which the bar density drops to zero.  The 
slanted line indicates the Sun--Galactic Center line for
the best-fit viewing angle.
The bar semimajor axis is $a = 3.6$~kpc.
\label{fig-radlagmosaic}}
(a) Model 2, $R_L$ = 4.0 kpc.  Shock lead angle $\approx 0\deg$,
bar viewing angle $31\deg$. \\
(b) Model 1, $R_L$ = 5.0 kpc.  Shock lead angle $\approx 5\deg$,
bar viewing angle $34\deg$. \\
(c) Model 5, $R_L$ = 6.0 kpc.  Shock lead angle $\approx 10\deg$,
bar viewing angle $20\deg$. \\
(d) Model 8, $R_L$ = 7.0 kpc.  Shock lead angle $\approx 15\deg$,
bar viewing angle $15\deg$. \\
(e) Model 4, $R_L$ = 8.0 kpc.  Shock lead angle $\approx 25\deg$,
bar viewing angle $8\deg$. 

\end{figure*}

We created a sequence of models including Model 1 to test 
the effect of varying the Lagrange radius or,
equivalently, the pattern speed of the bar.  
The sequence in Lagrange radius $R_L = 4$,
5, 6, 7, and 8 kpc yielded models 2, 1, 5, 8, and 4 respectively.
This sequence includes most of the best-fitting models 
(Model 3 is closely related).\footnote{
The groups of models \{33, 26, 22, 6\}, \{23, 12, 9\}, \{19, 13\}, 
and \{21, 18\} also comprise sets
where only the Lagrange radius is changed -- the effects are similar.}
Figure \ref{fig-radlagmosaic} shows face-on views of the gas density
in this sequence of models, like that of Figure \ref{fig-gasdensell}
for Model 1; Figure \ref{fig-radlaglv} shows \lv plots for
Models 2, 5, 8, and 4, to be compared with Figure \ref{fig-bestlv}.

The streamlines in Model 1 are not symmetric about the bar major axis;
in fact the major axis of the streamlines is rotated by about 5\deg\ 
with respect to it, the ``lead angle'' referred to in 
Section~\ref{sec-results1}.

\placefigure{fig-radlagmosaic}

Figure \ref{fig-radlagmosaic} shows that
the lead angle increases with the Lagrange radius, as far as 25\deg\
for the slowest bar.  The somewhat surprising result that several
models with grossly different Lagrange radii and lead angles all
appear to fit the \lv data reasonably well arises because the models
simply compensate by moving the best-fit viewing angle synchronously
with the changes in the lead angle.  The best-fit viewing angle stays
roughly constant with respect to the {\it shocks}, which means that it
also changes in a clockwise sense with respect to the bar, causing
\philsr\ to decrease.  Thus changes in viewing angle are strongly
coupled to the angle the gas streamlines make with the bar.

The systematic change in the location of the shocks has a relatively
simple explanation.  As the Lagrange radius is increased, the bar
pattern speed slows (for $R_L = 4.0$, 5.0, 6.0, 7.0, 8.0,
$\Omega_p = 54.2$, 41.9, 34.9, 30.2, 26.6
\kmskpc\ respectively).  Inside the Lagrange or corotation radius,
gas overtakes the gravitational potential well of the bar; the shocks are
caused as the gas climbs out of the well, slows down, and piles up
(Prendergast 1983).  
Although the shape of the gas streamlines is dependent on the full
gas-dynamics, the magnitude of the velocity for gas at a given radius
is roughly set by the gravitational acceleration from the mass
interior to it, which is the same in all five models.  In a frame
co-rotating with the bar, if the bar is slower, the gas is moving
faster as it overtakes the bar, so it climbs farther out of the
potential well before the shock pile-up occurs.  Therefore, in slower
pattern speed models, the shocks are farther ahead of the bar, in the
sense of more positive lead angle.  The increased speed of the gas
relative to the potential also increases the strength of the shocks.

The behavior of the shocks rules out slow bars, if we demand that the
Milky Way bar should resemble bars in other galaxies.  In external
galaxies, the prominent dust lanes frequently seen along the bar run
along the ``leading'' sides of the bar; the morphology of these dust
lanes and exemplary galaxies are discussed by Athana\-ssoula (1992b).
Strong bars generally have straight dust lanes while weaker bars
sometimes have curved dust lanes; in both cases, the dust lanes are
generally parallel to the bar, as in the shocks of Model 1, or angled 
slightly  in the sense of smaller lead angle.  These dust lanes are identified
with the high-density shocks, such as those in Figure
\ref{fig-gasdensell}, as discussed above. We know of no barred
galaxies that have dust lanes with a lead angle of more than a few
degrees; Athana\-ssoula (1992b) argued that therefore strong bars rotate
quickly.  Merrifield \& Kuijken (1995) have also
showed that the bar in NGC 936 rotates quickly, via a completely
independent method.

The position of the shocks in models 4 and 8 ($R_L = 7.0$ and 8.0 kpc), 
and in all
other slow bar models we have run, is grossly inconsistent with what
we know about barred galaxies.  We reject these models for this
reason, even though some of them formally fit the \lv diagram well.

\subsection{Bar strength and shape}
\label{sec-barshape}

The streamline plot of Figure \ref{fig-streamlines} and the contours
of observed velocity shown in Figure \ref{fig-velcontour} suggest that
the bar has to be strong and fairly elongated.  Only a massive bar can
produce the large non-circular motions needed to put gas at the
forbidden velocities observed in the \lv diagram.  If the gas
streamlines are less elongated than those seen in
Figure~\ref{fig-streamlines}, the regions with forbidden velocities
are smaller and subtend a smaller range of Galactic longitude.

\begin{figure*}[htb]

\psfig{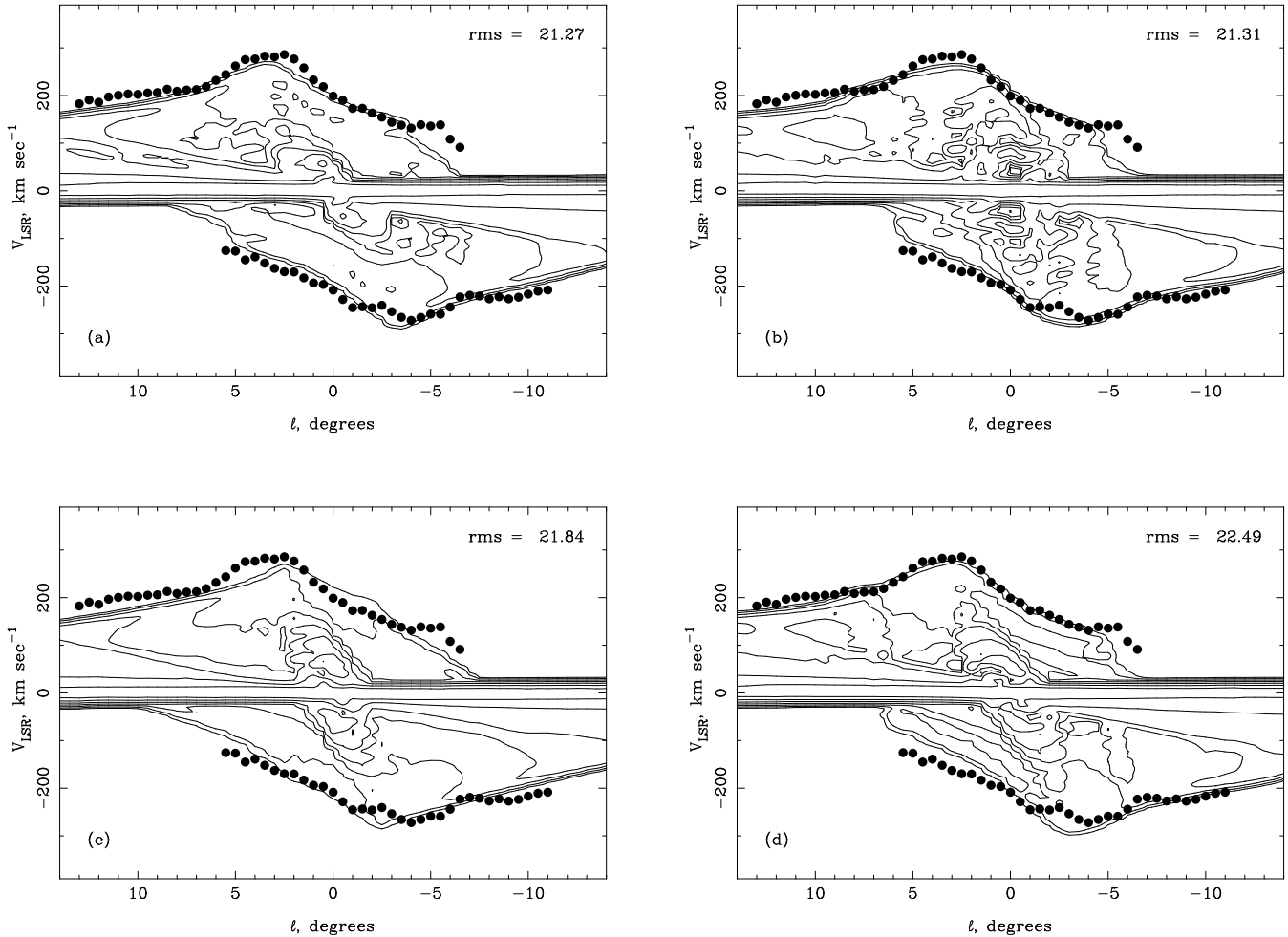}

\caption{Longitude-velocity diagrams for several models
with smaller bars of lower mass than that of Model 1.  
These models have $M_{bar}$ = 80\% that of Model 1, and $M_{bul}$
between 75\% and 125\% that of Model 1.
The lowest contour is 1.7 \lvunit\ of total gas and the contours
increase geometrically by a factor of 4.  There is a deficiency
of gas in the forbidden quadrants in these models.
\label{fig-weakbars}}
(a) Model 10, viewing angle 27\deg. \\
(b) Model 11, viewing angle 35\deg. \\
(c) Model 13, viewing angle 19\deg. \\
(d) Model 14, viewing angle 38\deg.

\end{figure*}

\placefigure{fig-weakbars}

In our best model, the bar component has a mass of
$M_{\rm bar} = 9.8~\times~10^9$~\msun, and the mass of the bulge 
component (within 1 kpc radius) is $M_{\rm bul} = 5.4~\times~10^9$~\msun.  

The effect of a weaker bar on
the \lv plots is shown in Figure~\ref{fig-weakbars}.
Models 10, 11, 13, and 14 show a significant deficit of gas in the 
forbidden quadrants, notably at $\ell \sim -5\deg$.
These models have smaller bars than Model 1 with lower
$M_{bar}$ (even though the bar density $\rho_{0,bar}$ is
somewhat higher).
Models 6, 9, and 12, which also have less massive bars
than Model 1, do somewhat
better at producing material in the forbidden quadrants, 
but only because the weaker forcing potential is partly
compensated for by the stronger shocks that occur in a slow-rotating
bar, as noted above.  However, models 6, 9, and 12, like all the other 
slow-bar models, have shocks in an implausible position and are not 
viable models for the Galaxy.

The bar must also be strong in the sense of having a large axis ratio.
The formal axis ratio of the Ferrers bar in Model 1 is 4:1, although
the actual axis ratio of the total mass distribution, when the bulge
and disk are included, is closer to 3:1 (cf.\ Figure ~\ref{fig-surfmass}).

\begin{figure*}[htb]

\psfig{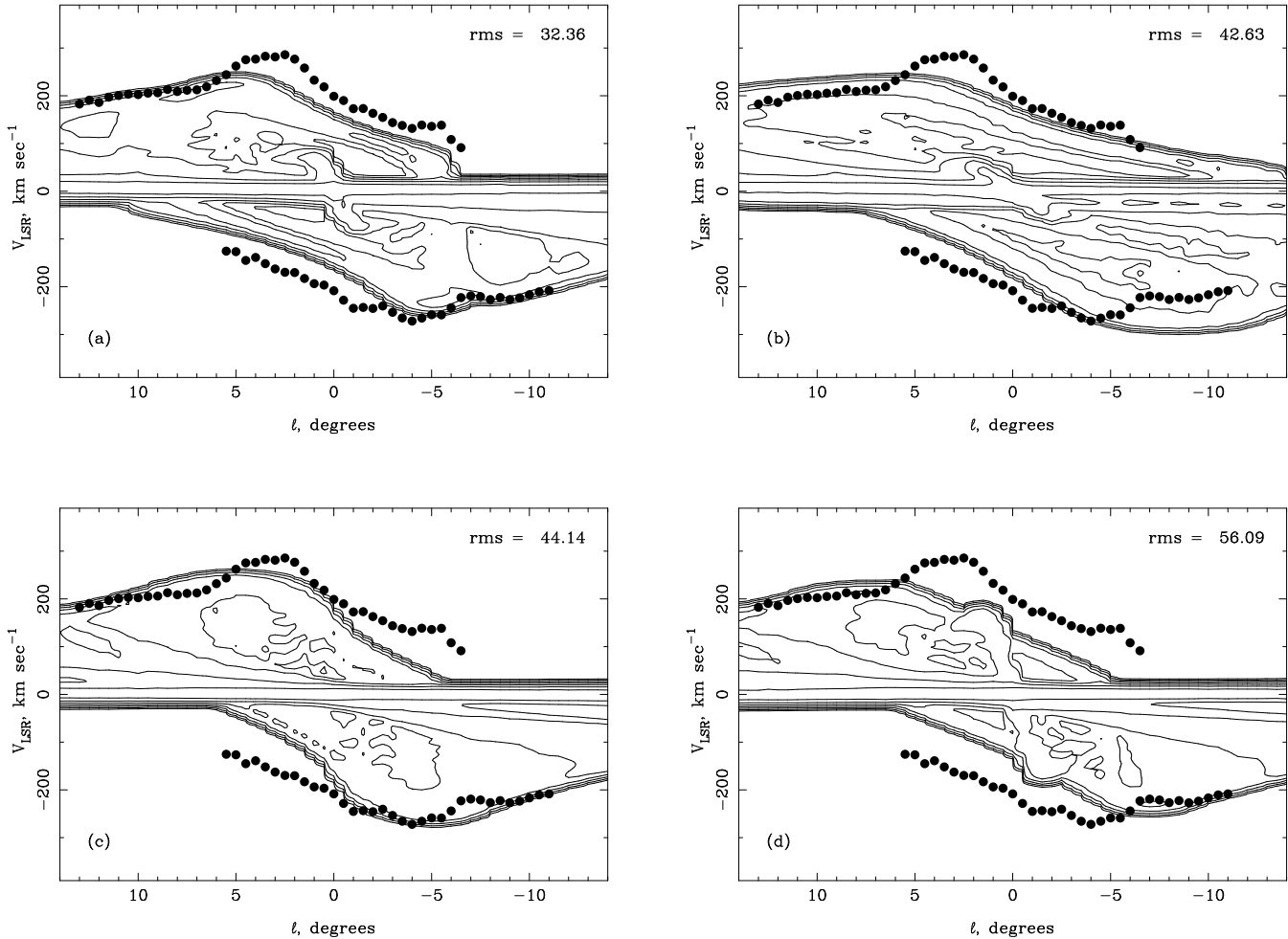}

\caption{Longitude-velocity diagrams for several models
with bars that have lower axis ratios than Model 1.
The lowest contour is 1.7 \lvunit\ of total gas and the contours
increase geometrically by a factor of 4.  These models have
EVCs whose peaks are not as sharp or as high as those observed.
\label{fig-fatbars}}
(a) Model 29, $a$ = 3.6 kpc, $b$ = 1.2 kpc, $a/b$ = 3.0,
viewing angle 27\deg. \\
(b) Model 37, $a$ = 4.0 kpc, $b$ = 1.3 kpc, $a/b$ = 3.1,
viewing angle 55\deg. \\
(c) Model 39, $a$ = 2.5 kpc, $b$ = 1.0 kpc, $a/b$ = 2.5,
viewing angle 35\deg. \\
(d) Model 46, $a$ = 3.0 kpc, $b$ = 1.3 kpc, $a/b$ = 2.3,
viewing angle 32\deg. 

\end{figure*}

\placefigure{fig-fatbars}

Models with smaller axis ratios generally do not reproduce the data
well.  Figure~\ref{fig-fatbars} shows \lv diagrams for several
models whose bar components have axis ratios smaller than that
of Model 1, with $a:b$ from 3.1:1 to 2.3:1.  The axis ratios of 
the total mass distributions are fatter still.  
Although these models have different
bar lengths, their appearance in \lv diagrams is similar:
they produce EVCs that are gently sloped, not
sharply peaked as seen in the observations.  In particular, the
decline of the EVC away from the peaks is fairly sharp in the
observations, but much too gentle in the models with low axis ratio
bars.

Even an extremely centrally concentrated model but wide-barred
potential such as
Model 46, which has a small dense bulge, does not produce sharp
peaks.  More massive
bars -- longer, more dense, or both -- do not successfully 
produce sharper peaks or better models: the most massive bars 
are models 45, 42, 37, 27, 43 and 44 ($M_{bar}$ = 41.2, 27.2, 22.7,
17.5, 15.9, and 15.9 $\times 10^9 \msun$, respectively).

As discussed above, the peaks in the \lv diagram are produced by the
strongly non-circular motions inside the bar while the steep decline
in the EVC as $|\ell|$ increases further is linked to the weakening of
the non-circular motions as the quadrupole field decays quickly with
Galactocentric distance.  An axisymmetric model with an unusual mass
distribution could be made to produce this behavior but could not, of
course, give rise to forbidden velocities.  A strong and elongated bar
is favored to produce both forbidden velocities and the narrow peaks
in the EVC.

\subsection{The presence of an inner Lindblad resonance}
\label{sec-ilr}

As already noted, the peaks of the \lv diagram arise from orbits just
outside the oval of high density gas in the center, where the
streamlines rotate to be highly angled to the bar rather than closely
aligned with it.  This rotation of the streamlines is related to the
presence of an inner Lindblad resonance (ILR) (Athana\-ssoula 1992a,b).
Bars with an ILR have a family of stellar orbits near the center that
are elongated perpendicular to the bar rather than along it, and the
rotated streamlines are related to these orbits.  Bars without an ILR
have only streamlines elongated along the bar; these streamlines would
yield peak gas velocities as they pass the center (Athana\-ssoula
1992b).

\begin{figure*}[htb]

\psfig{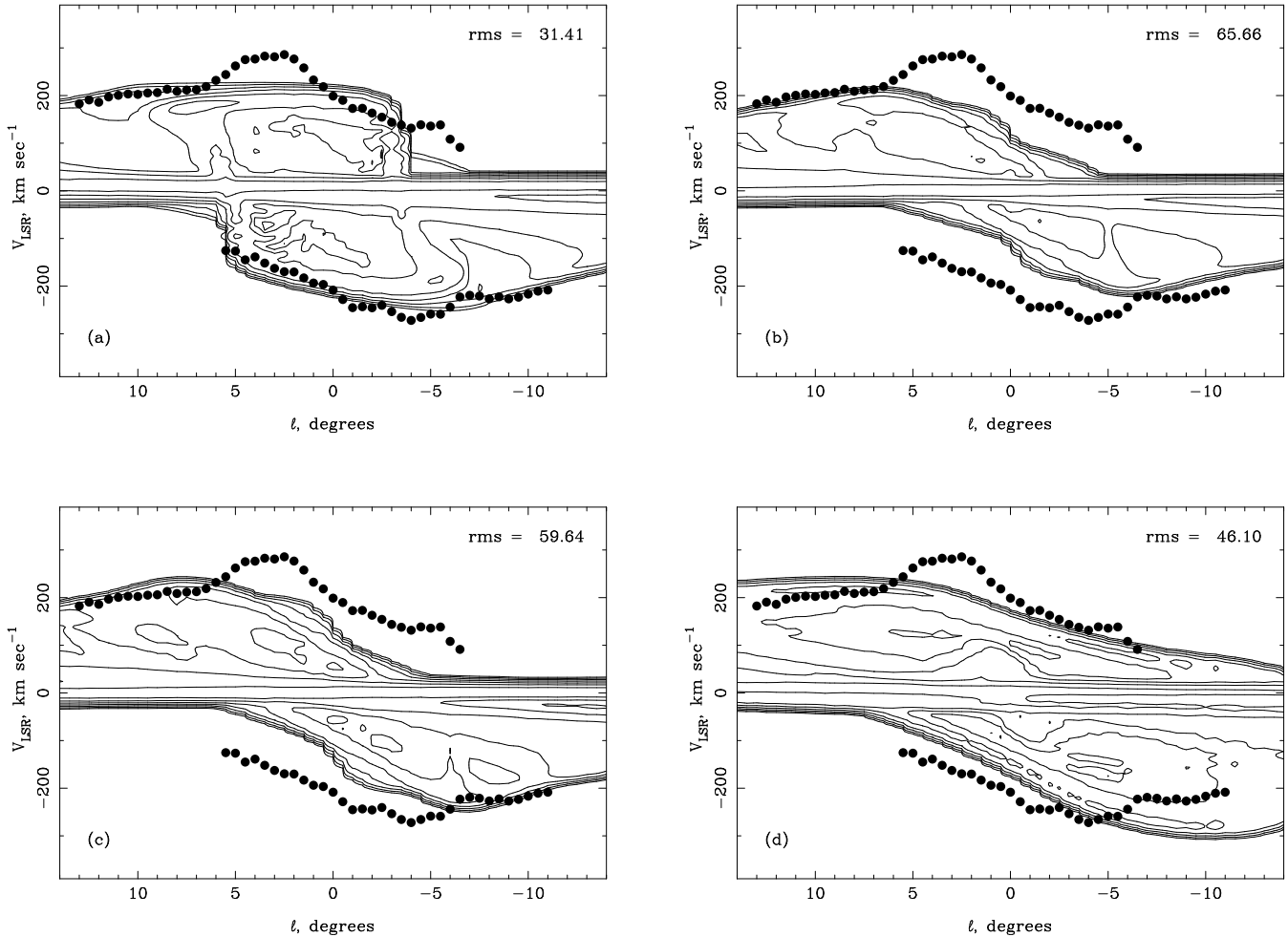}

\caption{Longitude-velocity diagrams for several models
which are less centrally concentrated than Model 1.
The lowest contour is 1.7 \lvunit\ of total gas and the contours
increase geometrically by a factor of 4.  These models exhibit
gently changing EVCs lacking the observed peaks.
\label{fig-noilrbars}}
(a) Model 27, $M_{bul}/M_{bar} = 0.019$, viewing angle 20\deg. \\
(b) Model 50, $M_{bul}/M_{bar} = 0.13$, viewing angle 20\deg. \\
(c) Model 48, $M_{bul}/M_{bar} = 0.16$, viewing angle 23\deg. \\
(d) Model 42, $M_{bul}/M_{bar} = 0.17$, viewing angle 55\deg. 

\end{figure*}

\placefigure{fig-noilrbars}

An inner Lindblad resonance forces the elongated orbits away from the
center, causing the highest bar-induced streaming 
velocities to occur some distance out, and
producing sharply defined peaks in the \lv diagram that are several
degrees apart.  If the bar did not have an ILR, the EVC peaks are not
necessarily as sharply defined, nor can they be separated by several
degrees in longitude, as is observed.  Model 27 is the least centrally
concentrated of all our models, and its best-fit \lv diagram has EVCs
without dominant peaks; the positive velocity EVC is nearly flat from
$\ell=-3\deg$ to $\ell=10\deg$.
The least centrally concentrated potentials are
models 27, 51, 45, 50, 48, and 42 ($M_{bul} / M_{bar}$ = 0.019,
0.096, 0.11, 0.13, 0.16, and 0.17 respectively).
Figure~\ref{fig-noilrbars} shows \lv diagrams for four of these
weakly-concentrated potentials, which have EVCs with weak
or gentle peaks.

The observed strength and separation of the peaks in the \lv diagram
suggests that the Galactic bar must have an ILR.  The central mass
concentration, represented in our model by the ``bulge'' component, is
responsible for the ILR, and is also necessary to cause the sharply
rising peaks in the EVC.

Our adopted modified Hubble profile for the central mass component has
a uniform density core, whereas the luminosity density in the Milky Way
rises all the way to the center as the $\sim -1.8$ power of the radius
(Becklin \& Neugebauer 1968).  The finite resolution of the grid
code vitiates attempts to simulate the effects of a
central cusp; strong gradients in the angular velocity
on scales below a few grid cells cannot be accurately
represented.
However, the small core radius in our best model, $r_c=0.2$ kpc (four
simulation grid cells), is well inside the ILR feature at
$R \sim 0.4$~kpc.  The existence of the ILR implied by the
EVC peaks requires only a concentrated mass within that radius,
so our conclusion is little affected by the details of the 
density profile.

\section{Discussion}

We have shown that gas flow in a barred model of the Galaxy can fit
many of the observed features of the \hi\ \lv diagram, most notably
the emission in the forbidden quadrants and the sharp peaks in velocity. 
Our best fit model was arrived at through adjusting the free parameters by
trial and error.  Although the model has been tuned, the number of
parameters is relatively small for a model of the Galactic potential.
Furthermore, the complexity of the dynamics governing the gas response
to the potential makes constructing a reasonably good model a
non-trivial pursuit.  This model does show that it is possible, and
that the reservations of Jenkins \& Binney (1994) regarding the
ability of simple gas-dynamical models to reproduce the data are
perhaps too pessimistic.

Our preferred model has a bar semi-major axis of 3.6 kpc.  The bar
component itself has an axis ratio of 4:1, although the ``bulge'' in
our model should also be considered as part of the bar, and the axis
ratio of the bar+bulge is somewhat fatter, approximately 3:1.  
The bar in this model rotates quickly, with a Lagrange radius of 5.0 kpc
(bar pattern speed 42 \kmskpc)
and the bar major axis is inclined at 34\deg\ to our line of sight.

Our model 1 differs in a number of important respects from the models
of the Milky Way bar proposed by Binney \etal\ (1991), and further
developed by Jenkins \& Binney (1994) and by Englmaier \& Gerhard (1998).
These authors favor a considerably smaller bar
and a higher pattern speed, placing corotation at
$R\sim 3.5$~kpc, because they employ a cusped $x_1$ orbit to give the
narrow peaks at $\ell \simeq \pm 3\deg$.
The $x_2$ orbit family, which
is much less extensive in their models than in ours, gives a smaller peak
very close to $\ell=0$.  While their models were developed to interpret
the CO \lv diagram, they fail to account for the large and extensive
forbidden velocities seen in \hi. 

The strength and size of the bar are required by forbidden velocities
in excess of 100 \kms\ extending as far as $\el = \pm 6\deg$.  Were
the true viewing angle much less than our preferred 34\deg, as favored
in some studies, the bar would have to be considerably longer to 
produce the observed forbidden velocities.  Our constraint on viewing
angle is not independent of the bar pattern speed, however, since
slower bars give better fits when viewed at smaller angles.
Models with a Lagrange radius of $R_L$ = 4.0 to 6.0 kpc 
(bar pattern speed $\Omega_p$ = 54 to 35 \kmskpc), i.e.\
fast-rotating bars, are favored; models with higher $R_L$ 
(lower $\Omega_p$) have shock patterns in the gas that
differ drastically from those observed in other barred galaxies.
It is unlikely that the viewing angle could be forced below 25\deg\
to the bar major axis.

We interpret the narrow velocity peaks at $\ell \simeq \pm 3\deg$ as
the signature of gas streaming along the bar past a nuclear ring in
the Milky Way which lies close to the location of the inner Lindblad
resonance.  If the bar is strong, the high speed of these streams does
not require an unusual radial mass profile --  the flow patterns
shown in Figures \ref{fig-axisymlv} \& \ref{fig-bestlv}
arise from two mass distributions that both,
when azimuthally averaged, give the circular velocity curve
shown in Figure~\ref{fig-axirotcurv}.  The mass distribution in the
inner Galaxy does have to be sufficiently concentrated for an ILR to
be present, however; if this were not the case, the peaks would lie
much closer to $\ell=0$.

The location of the peaks at $\ell \simeq \pm 3\deg$ requires the
semi-major axis of the nuclear ring to be $\sim 400$ pc -- on the small end
of the distribution of nuclear rings
seen in other barred galaxies (Buta \& Crocker 1993).  As nuclear rings in
external galaxies are generally highly gas rich (Helfer \& Blitz 1995;
Sofue 1996; Rubin, Kenney \& Young 1997), it is
no surprise that the associated velocity peaks in the Milky Way stand out
in CO as well as \hi.

We note that the rotation curve of our preferred model, shown in
Figure \ref{fig-axirotcurv}, indicates that the bulge and bar
components together dominate the rotation curve in the inner few kpc
of the Galaxy.  We cannot isolate the contribution of the dark halo
component, since our analytical model lumps the dark halo and the
axisymmetric part of the disk together.  However, since the Galaxy
{\it does} have a disk, it is clear that the dark halo cannot be very
dominant in this model.  Although this potential is not a unique model of
the Galaxy, as discussed above in Section \ref{sec-results1}, we
believe that any model that fits the \lv diagram will have to have
non-axisymmetric motions as strong as those in Model 1
and, hence, a bulge+bar which
dominates the rotation curve in the inner part of the Galaxy.
Englmaier \& Gerhard (1998) modeled the gas flow in the
inner Galaxy, using models derived from COBE photometry.
They found that the luminous matter must dominate over dark 
matter inside the solar circle, in order to match the terminal
velocity curve in the non-forbidden quadrants.

We do not claim that because our model 1 gives a reasonable fit, the
mass distribution in the inner Galaxy must necessarily be very close
to the analytic form we have assumed.  The real mass distribution in
the inner Galaxy is undoubtedly more complex than our simple
analytical model.  A different form of mass distribution will
yield somewhat different results for the best-fitting model
parameters.  However, we believe that the real mass distribution
will resemble Model 1 in its chief details: the strength and size 
of the bar, presence of an ILR, and viewing angle which is
not too close to end-on.

We have not attempted to satisfy the
many other constraints on the shape of the inner Galaxy, such as COBE
photometry, simultaneously.  The model is broadly consistent
with some results, such as the bar viewing angle determined by the
IRAS point sources (Weinberg 1992), the
magnitude offset of red clump stars (Stanek \etal\ 1997),
and the distribution of OH/IR stars (Sevenster \etal 1999).
Fux (1999) has compared the appearance of arm features 
produced in a self-consistent model with 
features in the CO and \hi\ \lv diagrams; his preferred model
has a bar of similar length, with an ILR, and which rotates
quickly, but the preferred viewing angle is somewhat smaller,
25\deg, and the bar is fatter.  Fux's comparison of models
to data emphasizes high-density gas, while ours probes mostly
low-density gas, which may be responsible for some of the 
differences.  The viewing angles in Fux's best model and in ours
are both incompatible with models which invoke a fairly 
end-on bar to account for the high microlensing 
optical depth towards the Galactic Bulge (Zhao \&
Mao 1996; see also Fux 1997).

The \lv diagrams synthesized from fluid models are sensitive to the
details of the potential and the viewing angle, and the comparison
with the data is unaffected by extinction.  For these reasons we
believe that the technique has great power to discriminate among
candidate models of the inner Milky Way.  We may eventually hope to
identify a model of the Galactic bar that satisfies photometric
constraints and fits both the CO and \hi\ kinematic data.

\acknowledgments


We are grateful to Dick van Albada and Lia Athana\-ssoula for providing
us with the gas dynamics code and for helpful comments and advice on
its use, to Harvey Liszt for providing the data of Liszt \& Burton
(1980) in electronic form, and to an anonymous referee for a thoughtful 
report.  This work was supported by NSF grant AST 96/17088 and NASA 
LTSA grant NAG 5-6037.  BJW acknowledges support from a Carnegie
postdoctoral fellowship.


\cp

\def\cp{\relax}

\cp


\input table1bare.tex


\cp

\cp

\cp

\cp

\cp

\cp

\cp

\cp

\cp

\cp

\cp

\cp

\cp

\cp





%



\cp

\end{document}

%% file: table1bare.tex



\begin{deluxetable}{lcccccccccc}
\scriptsize
\tablenum{1}
\tablecaption{Parameters of Models -- Ordered Best-fit to Worst-fit}
\label{table-modelpars}
\tablehead{
     &     \multicolumn{2}{c}{Bulge}  &   \multicolumn{2}{c}{Disk}   &   
     \multicolumn{3}{c}{Bar}   &  Lagrange   &  Best--fit   &  RMS velocity  \\
 Model  & $\rho_{0,{\rm bul}}$, & $r_c$, &
    $\Sigma_0$, &  $R_c$, &  $\rho_{0, {\rm bar}}$, &
    $a$, &  $b$, &  radius & viewing angle & deviation,  \\
 No.  & \mden & kpc & \msurfden  & kpc & \mden &
     kpc &  kpc  & $R_L$, kpc & \philsr &  \kms 
}
\startdata
1  &    40  &  0.2  &   571  &  3.5  &  2.0  &  3.6  &  0.9  &  5.0  &  34\deg  &   16.54  \nl
2  &    40  &  0.2  &   571  &  3.5  &  2.0  &  3.6  &  0.9  &  4.0  &  32\deg  &   17.30  \nl
3  &    30  &  0.2  &   571  &  3.5  &  2.0  &  3.6  &  0.9  &  4.5  &  33\deg  &   17.90  \nl
4  &    40  &  0.2  &   571  &  3.5  &  2.0  &  3.6  &  0.9  &  8.0  &   8\deg  &   19.68  \nl
5  &    40  &  0.2  &   571  &  3.5  &  2.0  &  3.6  &  0.9  &  6.0  &  20\deg  &   19.82  \nl
6  &    30  &  0.2  &   571  &  3.5  &  2.5  &  2.5  &  0.83  & 8.0  &  73\deg  &   20.47  \nl
7  &    40  &  0.2  &   543  &  3.5  &  3.44  & 3.0  &  0.8   & 3.5  &  36\deg  &   20.82  \nl
8  &    40  &  0.2  &   571  &  3.5  &  2.0   & 3.6  &  0.9   & 7.0  &  14\deg  &   20.90  \nl
9  &    50  &  0.15  &  570  &  3.5  &  2.75  & 3.0  &  0.75  & 8.0  &  85\deg  &   21.21  \nl
10 &    70  &  0.15  &  571  &  3.5  &  2.75  & 3.0  &  0.75  & 3.6  &  27\deg  &   21.27  \nl
11 &    40  &  0.2  &   571  &  3.5  &  2.75  & 3.0  &  0.75  & 3.6  &  35\deg  &   21.31  \nl
12 &    50  &  0.15  &  570  &  3.5  &  2.75  & 3.0  &  0.75  & 7.2  &  91\deg  &   21.72  \nl
13 &    30  &  0.2  &   571  &  3.5  &  2.75  & 3.0  &  0.75  & 5.0  &  19\deg  &   21.84  \nl
14 &    50  &  0.2  &   571  &  3.5  &  2.75  & 3.0  &  0.75  & 3.6  &  38\deg  &   22.48  \nl
15 &    60  &  0.15  &  571  &  3.5  &  2.0   & 3.6  &  0.9   & 5.0  &  20\deg  &   22.49  \nl
16 &    70  &  0.15  &  570  &  3.5  &  2.75  & 3.0  &  0.75  & 3.6  &  26\deg  &   22.55  \nl
17 &    60  &  0.15  &  570  &  3.5  &  2.75  & 3.0  &  0.75  & 3.6  &  30\deg  &   22.64  \nl
18 &    65  &  0.15  &  570  &  3.5  &  2.75  & 3.0  &  0.75  & 8.0  &  79\deg  &   22.73\tablenotemark{a} \nl
19 &    30  &  0.2  &   571  &  3.5  &  2.75  & 3.0  &  0.75  & 3.6  &  28\deg  &   23.27  \nl
20 &    80  &  0.15  &  570  &  3.5  &  2.75  & 3.0  &  0.75  & 3.6  &  25\deg  &   23.79  \nl
21 &    65  &  0.15  &  570  &  3.5  &  2.75  & 3.0  &  0.75  & 3.6  &  23\deg  &   25.60  \nl
22 &    30  &  0.2  &   571  &  3.5  &  2.5  &  2.5  &  0.83  & 5.0  &  14\deg  &   26.86\tablenotemark{b}  \nl
23 &    50  &  0.15  &  571  &  3.5  &  2.75  & 3.0  &  0.75  & 3.6  &  21\deg  &   28.44  \nl
24 &    40  &  0.2  &   543  &  3.5  &  2.2  &  3.0  &  1.0  &  3.5  &  32\deg  &   28.96  \nl
25 &    50  &  0.2  &   571  &  3.5  &  2.0  &  3.6  &  0.9  &  5.0  &  23\deg  &   29.07  \nl
26 &    30  &  0.2  &   571  &  3.5  &  2.5  &  2.5  &  0.83  & 4.0  &  29\deg  &   29.18  \nl
27 &     5  &  0.15  &  570  &  3.5  &  2.75  & 3.8  &  1.0  &  4.4  &  20\deg  &   31.41  \nl
28 &    40  &  0.2  &   571  &  3.5  &  1.8  &  3.0  &  1.0  &  3.5  &  31\deg  &   32.23  \nl
29 &    60  &  0.15  &  571  &  3.5  &  1.4  &  3.6  &  1.2  &  4.2  &  27\deg  &   32.36  \nl
30 &    50  &  0.15  &  686  &  3.5  &  2.75  & 3.0  &  0.75  & 3.6  &  29\deg  &   33.21  \nl
31 &    40  &  0.2  &   571  &  3.5  &  2.5  &  2.5  &  0.83  & 3.0  &  31\deg  &   33.34  \nl
32 &    100 &  0.15  &  570  &  3.5  &  2.75  & 3.0  &  0.75  & 3.6  &  17\deg  &   36.29  \nl
33 &    30  &  0.2  &   571  &  3.5  &  2.5  &  2.5  &  0.83  & 3.0  &  31\deg  &   37.91  \nl
34 &    80  &  0.15  &  571  &  3.5  &  1.4  &  3.6  &  1.2  &  4.2  &  27\deg  &   38.45  \nl
35 &    40  &  0.15  &  571  &  3.5  &  1.4  &  3.6  &  1.2  &  4.2  &  23\deg  &   39.49  \nl
36 &    25  &  0.15  &  571  &  3.5  &  2.75  & 3.0  &  0.75  & 3.6  &  25\deg  &   41.94  \nl
37 &    50  &  0.2  &   700  &  2.0  &  2.0  &  4.0  &  1.3  &  5.0  &  55\deg  &   42.63  \nl
38 &    40  &  0.2  &   543  &  3.5  &  1.53  & 3.0  &  1.2  &  3.5  &  31\deg  &   43.97  \nl
39 &    40  &  0.2  &   571  &  3.5  &  2.5  &  2.5  &  1.0  &  3.0  &  35\deg  &   44.14  \nl
40 &    50  &  0.15  &  570  &  3.5  &  1.55  & 3.0  &  1.0  &  3.6  &  23\deg  &   44.20  \nl
41 &    130 &  0.15  &  570  &  3.5  &  2.75  & 3.0  &  0.75 &  3.6  &  23\deg  &   45.55  \nl
42 &    35  &  0.2  &   800  &  2.0  &  2.4  &  4.0  &  1.3  &  5.0  &  55\deg  &   46.10  \nl
43 &    40  &  0.2  &   750  &  2.0  &  2.2  &  3.0  &  1.2  &  3.5  &  36\deg  &   49.09  \nl
44 &    40  &  0.25 &  467  &  3.0  &  2.2  &  3.0  &  1.2  &  3.5  &  49\deg  &   50.31  \nl
45 &    35  &  0.2  &  800  &  2.0  &  2.4  &  4.0  &  1.6  &  5.0  &  57\deg  &   54.66  \nl
46 &    200  &  0.1  &  567  &  3.0  &  1.5  &  3.0  &  1.3  &  3.5  &  32\deg  &   56.09  \nl
47 &    65  &  0.15  &  570  &  3.5  &  1.08 &  3.0  &  1.2  &  3.6  &  28\deg  &   58.08  \nl
48 &    35  &  0.15  &  600  &  3.0  &  1.8  &  3.0  &  1.3  &  3.5  &  23\deg  &   59.64  \nl
49 &    50  &  0.15  &  570  &  3.5  &  1.08 &  3.0  &  1.2  &  3.6  &  24\deg  &   62.94\tablenotemark{c}  \nl
50 &    25  &  0.15  &  567  &  3.0  &  1.5  &  3.0  &  1.3  &  3.5  &  20\deg  &   65.66  \nl
51 &    15  &  0.15  &  570  &  3.5  &  0.8  &  3.5  &  1.5  &  4.2  &  23\deg  &   86.97  \nl

\enddata
\tablenotetext{a}{RMS velocity deviation = 19.90 \kms\ at \philsr = --9\deg}
\tablenotetext{b}{RMS velocity deviation = 17.68 \kms\ at \philsr = 103\deg}
\tablenotetext{c}{RMS velocity deviation = 54.24 \kms\ at \philsr = --30\deg}
\end{deluxetable}
